\documentclass[12pt]{article}
\pdfoutput=1

\usepackage[a4paper,text={16.8cm,22.4cm}]{geometry}
\usepackage{amsmath,amsfonts,braket,slashed,amssymb,tikz,bm,psfrag,graphicx,color,dsfont,euscript}
\usepackage{multicol}
\usepackage{ctable}
\usepackage[small,labelfont=bf]{caption}
\usepackage{wrapfig}

\RequirePackage[sort&compress,square,comma,numbers]{natbib}
\allowdisplaybreaks
\newcommand{\A}{{\EuScript A}}
\newcommand{\X}{{\EuScript X}}
\newcommand{\G}{{\EuScript G}}

\newcommand{\nsl}{\rlap{\hspace{0.25mm}/}{n}}
\newcommand{\nbsl}{\rlap{\hspace{0.25mm}/}{\bar n}}
\newcommand{\Asl}{\rlap{\hspace{0.7mm}/}{\A}}
\newcommand{\Gsl}{\rlap{\hspace{0.2mm}/}{\G}}
\newcommand{\Dsl}{\rlap{\hspace{0.75mm}/}{D}}

\newcommand{\braces}[1]{[\hspace{-0.5mm}[ #1 ]\hspace{-0.5mm}]}

\begin{document}

\begin{titlepage}

\begin{flushright}
\normalsize
MITP/19-078\\ 
December 20, 2019
\end{flushright}

\vspace{1.0cm}
\begin{center}
\Large\bf
Factorization at Subleading Power and Endpoint-Divergent Convolutions 
in {\boldmath $h\to\gamma\gamma$} Decay 
\end{center}

\vspace{0.5cm}
\begin{center}
Ze Long Liu$^{a,b}$ and Matthias Neubert$^{a,c}$\\
\vspace{0.7cm} 
{\sl ${}^a$PRISMA$^+$ Cluster of Excellence \& Mainz Institute for Theoretical Physics\\
Johannes Gutenberg University, 55099 Mainz, Germany\\[3mm]
${}^b$Theoretical Division, Los Alamos National Laboratory, Los Alamos, NM 87545, U.S.A.\\[3mm]
${}^c$Department of Physics \& LEPP, Cornell University, Ithaca, NY 14853, U.S.A.}
\end{center}

\vspace{0.8cm}
\begin{abstract}
It is by now well known that, at subleading power in scale ratios, factorization theorems for high-energy cross sections and decay amplitudes contain endpoint-divergent convolution integrals. The presence of endpoint divergences hints at a violation of simple scale separation. At the technical level, they indicate an unexpected failure of dimensional regularization and the $\overline{\rm MS}$ subtraction scheme. In this paper we start a detailed discussion of factorization at subleading power within the framework of soft-collinear effective theory. As a concrete example, we factorize the decay amplitude for the radiative Higgs-boson decay $h\to \gamma\gamma$ mediated by a $b$-quark loop, for which endpoint-divergent convolution integrals require both dimensional and rapidity regulators. We derive a factorization theorem for the decay amplitude in terms of bare Wilson coefficients and operator matrix elements. We show that endpoint divergences caused by rapidity divergences cancel to all orders of perturbation theory, while endpoint divergences that are regularized dimensionally can be removed by rearranging the terms in the factorization theorem. We use our result to resum the leading double-logarithmic corrections of order $\alpha_s^n\ln^{2n+2}(-M_h^2/m_b^2)$ to the decay amplitude to all orders of perturbation theory.
\end{abstract}

\end{titlepage}

\tableofcontents
\vspace{8mm}

\section{Introduction}

Factorization theorems play an important role in understanding the structure of observables sensitive to several different mass or distance scales. They provide the means to separate long-distance from short-distance phenomena and resum large logarithmic corrections to all orders of perturbation theory. Whereas at leading power in scale ratios a typical factorization theorem consists of a product (often in the convolution sense) of functions each associated with a single scale, at subleading power several complications arise. Yet, in times when high-precision calculations of scattering and decay processes are state-of-the-art, going beyond the leading order in scale ratios has become a necessity in many cases. 

For example, obtaining more precise fixed-order calculations of collider cross sections has been an important goal for many years. A major difficulty in these calculations is the proper handling of the infrared singularities that arise in both virtual and real contributions. A method based on $N$-jettiness (${\cal T}_N$) slicing \cite{Boughezal:2015eha,Gaunt:2015pea} allows one to obtain the next-to-next-to-leading order (NNLO) QCD result from a much easier next-to-leading order (NLO) calculation, combined with information about the singular dependence of the cross section on the ${\cal T}_N$  resolution variable \cite{Stewart:2010tn}. Calculations of the dominant power corrections in ${\cal T}_N/Q$ have helped to improved the numerical stability for several processes \cite{Moult:2016fqy,Boughezal:2016zws,Ebert:2018lzn}. In other cases, the quantity of interest may itself be a power-suppressed process. An example is the decay of a heavy scalar particle into a pair of light fermions \cite{Alte:2018nbn,Alte:2019iug}.

Soft-collinear effective theory (SCET) \cite{Bauer:2000yr,Bauer:2001yt,Bauer:2002nz,Beneke:2002ph,Beneke:2002ni} provides a convenient framework for addressing the problem of factorization and scale separation using the powerful tools of effective field theory. Much recent work has focused on exploring the structure of factorization theorems at subleading order in power counting. In a first step, this involves the construction of a basis of subleading SCET operators (see e.g.\ \cite{Chay:2002vy,Pirjol:2002km,Hill:2004if,Beneke:2005gs,Moult:2017rpl,Feige:2017zci}) and the calculation of their anomalous dimensions \cite{Hill:2004if,Beneke:2005gs,Alte:2018nbn,Beneke:2017ztn,Beneke:2018rbh}. Specific applications that have been discussed include the threshold resummation at subleading power for the Drell-Yan and Higgs production processes \cite{Bonocore:2015esa,Bonocore:2016awd,Bahjat-Abbas:2018hpv,Beneke:2018gvs,Beneke:2019mua,Beneke:2019oqx}, the study of power corrections to event shapes \cite{Moult:2018jjd} and transverse-momentum distributions \cite{Ebert:2018gsn,Moult:2019vou}, and an analysis of power corrections to the exclusive $B$-meson decay $B\to\gamma l\bar\nu_l$ \cite{Beneke:2018wjp}. One finds that such factorization theorems contain a sum over several products of Wilson coefficients times operator matrix elements, where the relevant SCET operators mix under renormalization. At the same time new complications arise, which do not occur at leading power. The most puzzling one is the appearance of endpoint-divergent convolution integrals over products of component functions each depending on a single scale. Such endpoint divergences were first observed in the analysis of exclusive, non-leptonic and radiative weak decays of $B$ mesons in the context of QCD factorization \cite{Beneke:2000ry,Beneke:2001ev,Beneke:2001at,Beneke:2003zv}. While the corresponding decay amplitudes were shown to factorize at leading order in $\Lambda_{\rm QCD}/m_b$, some of the power-suppressed corrections were found to contain endpoint-divergent convolution integrals of hard-scattering kernels with meson light-cone distribution amplitudes. These divergences have been interpreted as a break-down of factorization in the non-perturbative domain. To the best of our knowledge, the first examples of endpoint-divergent convolution integrals in a {\em perturbative\/} setting have been found in the study of the exclusive decay $B\to\chi_{cJ} K$ \cite{Beneke:2008pi} and of non-local power corrections to the inclusive radiative decay $B\to X_s\gamma$ \cite{Benzke:2010js}. 

It is by now well known that endpoint-divergent convolution integrals are a generic feature of SCET factorization theorems beyond leading power. In simple cases, where the relevant momentum modes are hard, collinear and ultrasoft and the relevant effective theory is the so-called SCET-1, the endpoint divergences are regularized by the regulator $\epsilon=(4-D)/2$ of dimensional regularization. Nevertheless, these divergences indicate an unexpected failure of dimensional regularization and the $\overline{\rm MS}$ subtraction scheme, because the $1/\epsilon^n$ pole terms arising in integrals over products of functions depending on different mass scales cannot be removed by means of the renormalization factors of these individual functions. Hence, a simple scale separation is spoiled by these divergences. Standard tools of renormalization theory are then no longer sufficient to obtain a well-defined, renormalized factorization theorem involving convergent convolutions over renormalized functions. In more complicated cases, in which the relevant momentum modes are hard, collinear and soft and the relevant effective theory is the so-called SCET-2, one encounters the additional complication that there exist endpoint divergences which are not regularized by the dimensional regulator but require an additional rapidity regulator. 

In this work, we consider the $b$-quark induced amplitude for the decay of a Higgs boson into two photons, $h\to(b\bar b)^*\to\gamma\gamma$, as a concrete example of a quantity obeying a subleading-power factorization theorem. The power suppression arises from the need for an insertion of the $b$-quark mass, which implies a suppression by one power in the expansion parameter $\lambda=m_b/M_h\ll 1$. (The corresponding amplitude for a heavy vector boson decaying into two light particles would not feature such a suppression.) This power suppression introduces sensitivity to a kinematic region in which a soft quark is exchanged between the two photons. Contrary to soft gluons, soft quarks are power suppressed in SCET, and their emission cannot be described in terms of soft Wilson lines \cite{Moult:2019mog,Moult:2019uhz}. The factorized decay amplitude in SCET is subject to both types of endpoint divergences: those regularized with the dimensional regulator and those regularized by a rapidity regulator. This example therefore illustrates the most complicated situation in which endpoint divergences can arise.

The $b$-quark induced $h\to\gamma\gamma$ amplitude has been analyzed long ago using conventional tools of perturbative quantum field theory \cite{Kotsky:1997rq,Akhoury:2001mz}, and it was shown how to resum the leading double-logarithmic QCD corrections in the ratio $M_h^2/m_b^2$ to all orders of perturbation theory. More recently, the scope of this approach has been extended significantly by validating its validity up to three-loop order for several relevant processes \cite{Penin:2014msa,Liu:2017vkm,Liu:2018czl}. Here we present the first SCET analysis of the $b$-quark induced $h\to\gamma\gamma$ process. Our goal is to establish an all-order factorization theorem for the decay amplitude and develop a framework that will in principle allow for the resummation of large logarithms at arbitrary order in perturbation theory. This paper takes the first steps toward this goal. In Section~\ref{sec:2} we derive a factorization theorem in terms of products (in the convolution sense) of bare Wilson coefficients with matrix elements of bare SCET operators. In Sections~\ref{sec:3} and~\ref{sec:4} we explicitly calculate the matching coefficients and operator matrix elements at NLO in the QCD coupling $\alpha_s$, corresponding to the two-loop approximation for the decay amplitude. While formulating the proper procedure of $\overline{\rm MS}$-like renormalization in the presence of endpoint divergences is still an open challenge (see e.g.\ the recent works \cite{Alte:2018nbn,Moult:2019uhz,Beneke:2019oqx}), we take important steps in the direction of solving this problem. We introduce a suitable analytic regulator for the rapidity divergences in Section~\ref{sec:rapidity}, where we also derive the exact conditions required for the cancellation of rapidity divergences to all orders of perturbation theory. This cancellation is highly non-trivial in our case, since it arises between two terms in the factorization formula depending on different physical scales (hard and collinear versus hard, hard-collinear and soft). In Section~\ref{sec:subtr} we succeed to rearrange the factorization formula in such a way that all endpoint divergences -- those requiring analytic regulators and those regularized dimensionally -- are removed. Our main result, a factorization formula free of endpoint divergences, is presented in (\ref{fact4}). In Section~\ref{sec:6} we identify the leading double logarithms contributing to the decay amplitude and resum them to all orders of perturbation theory, carefully taking into account the time-like kinematics of the $h\to\gamma\gamma$ process. We thus resum logarithms of $M_h^2/m_b^2$ along with $i\pi$ terms \cite{Ahrens:2008qu,Ahrens:2008nc}, and we also include the scale evolution of $\alpha_s(\mu)$, thereby extending the results obtained in \cite{Kotsky:1997rq,Akhoury:2001mz,Penin:2014msa,Liu:2017vkm,Liu:2018czl} in several important ways. Finally, in Section~\ref{sec:7} we summarize our results and discuss some aspects that will be relevant for extending the resummation beyond the leading double-logarithmic approximation.

As this paper deals with a rather sophisticated application of scale factorization, we shall assume that the reader is familiar with the basic formalism of SCET as described, e.g., in the founding papers \cite{Bauer:2000yr,Bauer:2001yt,Bauer:2002nz,Beneke:2002ph,Beneke:2002ni}.

\section{Bare factorization theorem}
\label{sec:2}

In this paper, the $h\to\gamma\gamma$ amplitude is always to be understood as the contribution mediated by virtual $b$-quarks. We neglect the (numerically dominant) contribution from the top-quark loop, which does not give rise to the large logarithms investigated here. At NLO in QCD perturbation theory, the amplitude for the process $h\to(b\bar b)^*\to\gamma\gamma$ is given by \cite{Spira:1995rr} 
\begin{equation}\label{QCDresult}
\begin{aligned}
   {\cal M}_b(h\to\gamma\gamma)
   &= {\cal M}_0\,\Bigg\{ \left( \frac{L^2}{2} - 2 \right) 
    + \frac{C_F\alpha_s(\mu)}{4\pi}\,\bigg[ - \frac{L^4}{12} - L^3 + \left( 4 - \frac{2\pi^2}{3} \right) L^2 \\
   &\hspace{16mm}\mbox{}+ \left( 4 + \frac{2\pi^2}{3} + 16\zeta_3 \right) L - 36 + 4\zeta_3 - \frac{\pi^4}{5} \\
   &\hspace{16mm}\mbox{}- \left( 3L^2 - 6L - 12 \right) \ln\frac{m_b^2(\mu)}{\mu^2} \bigg] 
    + {\cal O}\bigg(\frac{m_b^2}{M_h^2}\bigg) \Bigg\} \,,
\end{aligned}
\end{equation}
where $L=\ln[-M_h^2/m_b^2(\mu)-i0]=\ln[M_h^2/m_b^2(\mu)]-i\pi$ is the large logarithm in the problem, and 
\begin{equation}\label{M0def}
   {\cal M}_0 = \frac{N_c\alpha_b}{\pi}\,\frac{y_b(\mu)}{\sqrt{2}}\,m_b(\mu)\,g_\perp^{\mu\nu}\,
    \varepsilon_\mu^*(k_1)\,\varepsilon_\nu^*(k_2) \,, \qquad
   g_\perp^{\mu\nu} = g^{\mu\nu} - \frac{k_1^\mu k_2^\nu+k_1^\nu k_2^\mu}{k_1\cdot k_2} \,.
\end{equation}
Here $\alpha_b\equiv e_b^2/(4\pi)$ (with $e_b=-e/3$) is the electromagnetic coupling of $b$-quarks. The running $b$-quark mass $m_b$, the Yukawa coupling $y_b$ and the QCD coupling $\alpha_s$ are defined in the $\overline{\rm MS}$ scheme. The question how to set the renormalization scale in these quantities has been a subject of some discussion (see e.g.\ \cite{Spira:1995rr}), and it is most pressing for the definition of $m_b$, which enters as an argument in the large logarithm $L$ and as a prefactor. The difference between the $b$-quark pole mass $m_b^{\rm pole}$ and the running mass $m_b(M_h)$ is almost a factor of~2, and hence scale ambiguities can have a large impact on the decay rate. This issue was particularly pressing in times when low-energy supersymmetry was still a viable option, because in such models the Higgs coupling to $b$-quarks could be enhanced by a large factor $\tan\beta\gg 1$, making it numerically comparable with the top-quark induced contribution, which is the leading contribution in the Standard Model (SM). But even in the context of the SM an accurate prediction for the $b$-quark induced contribution is desirable. The uncertainty associated with scale setting in the $b$-quark mass and Yukawa coupling is not unimportant in the context of otherwise very precise, higher-order calculations of the $h\to\gamma\gamma$ decay amplitude and the related $gg\to h$ production amplitude (see \cite{Spira:2016ztx} for a review).

In order to address this question rigorously, one should factorize the two hierarchical scales $M_h\gg m_b$ and resum the large logarithms $L$ to all orders of perturbation theory. This resummation has first been accomplished in \cite{Kotsky:1997rq,Akhoury:2001mz} using conventional tools of perturbative QCD, but only at the level of the leading double logarithms $\alpha_s^n\hat L^{2n+2}$, where $\hat L=\ln(M_h^2/m_b^2)$. It was found that 
\begin{equation}\label{Penin}
\begin{aligned}
   {\cal M}_b(h\to\gamma\gamma) \big|_{\rm LDL}
   &= {\cal M}_0\,\hat L^2\,\sum_{n=0}^\infty\,\frac{n!}{(2n+2)!} \left( 
    - \frac{C_F\alpha_s}{2\pi}\,\hat L^2 \right)^n \\
   &= {\cal M}_0\,\frac{\hat L^2}{2}\,
    {}_2F_2\bigg(1,1;\frac32,2;-\frac{C_F\alpha_s}{8\pi}\,\hat L^2\bigg) \,,
\end{aligned}
\end{equation}
where ${}_2F_2(\dots)$ is a generalized hypergeometric function. Note that in this approximation the question of the scale choice is not resolved, because the leading double-logarithmic (LDL) approximation is not a consistent approximation in the sense that it ignores terms that are parametrically larger than ${\cal O}(1)$.

Analyzing the QCD diagrams giving rise to the decay amplitude at one- and two-loop order using the method of regions \cite{Beneke:1997zp,Smirnov:1998vk,Smirnov:2002pj}, we find that the amplitude receives leading contributions from the following momentum regions of loop momenta:
\begin{equation}
\begin{aligned}
   \mbox{hard ($h$)}: &\qquad \ell^\mu\sim M_h\,(1,1,1) \\
   \mbox{$n_1$-collinear ($c$)}: &\qquad \ell^\mu\sim M_h\,(\lambda^2,1,\lambda) \\
   \mbox{$n_2$-collinear ($\bar c$)}: &\qquad \ell^\mu\sim M_h\,(1,\lambda^2,\lambda) \\
   \mbox{soft ($s$)}: &\qquad \ell^\mu\sim M_h\,(\lambda,\lambda,\lambda)
\end{aligned}
\end{equation}
We use a power counting where $\lambda=m_b/M_h$ is the small parameter in the construction of the effective field theory. When soft exchanges are present, one also needs regions with hard-collinear scaling:
\begin{equation}\label{hcfields}
\begin{aligned}
   \mbox{$n_1$-hard-collinear ($hc$)}: &\qquad \ell^\mu\sim M_h\,(\lambda,1,\lambda^{\frac32}) \\
   \mbox{$n_2$-hard-collinear ($\overline{hc}$)}: &\qquad \ell^\mu\sim M_h\,(1,\lambda,\lambda^{\frac32})
\end{aligned}
\end{equation}
Here $n_1^\mu$ and $n_2^\mu$ are light-like momenta aligned with the directions $k_1^\mu$ and $k_2^\mu$ of the external photons, respectively, and satisfying $n_1\cdot n_2=2$. In the rest frame of the Higgs boson, we can choose $n_1^\mu=(1,0,0,1)$ and $n_2^\mu=(1,0,0,-1)$. Above we have indicated the scalings of the components $(n_1\cdot\ell,n_2\cdot\ell,\ell_\perp)$ of the loop momentum in the decomposition
\begin{equation}
   \ell^\mu = (n_1\cdot\ell)\,\frac{n_2^\mu}{2} + (n_2\cdot\ell)\,\frac{n_1^\mu}{2} + \ell_\perp^\mu \,.
\end{equation}
It will be useful below to introduce also the conjugate vectors $\bar n_1^\mu\equiv n_2^\mu$ and $\bar n_2^\mu\equiv n_1^\mu$.

\begin{figure}[t]
\begin{center}
\includegraphics[width=0.55\textwidth]{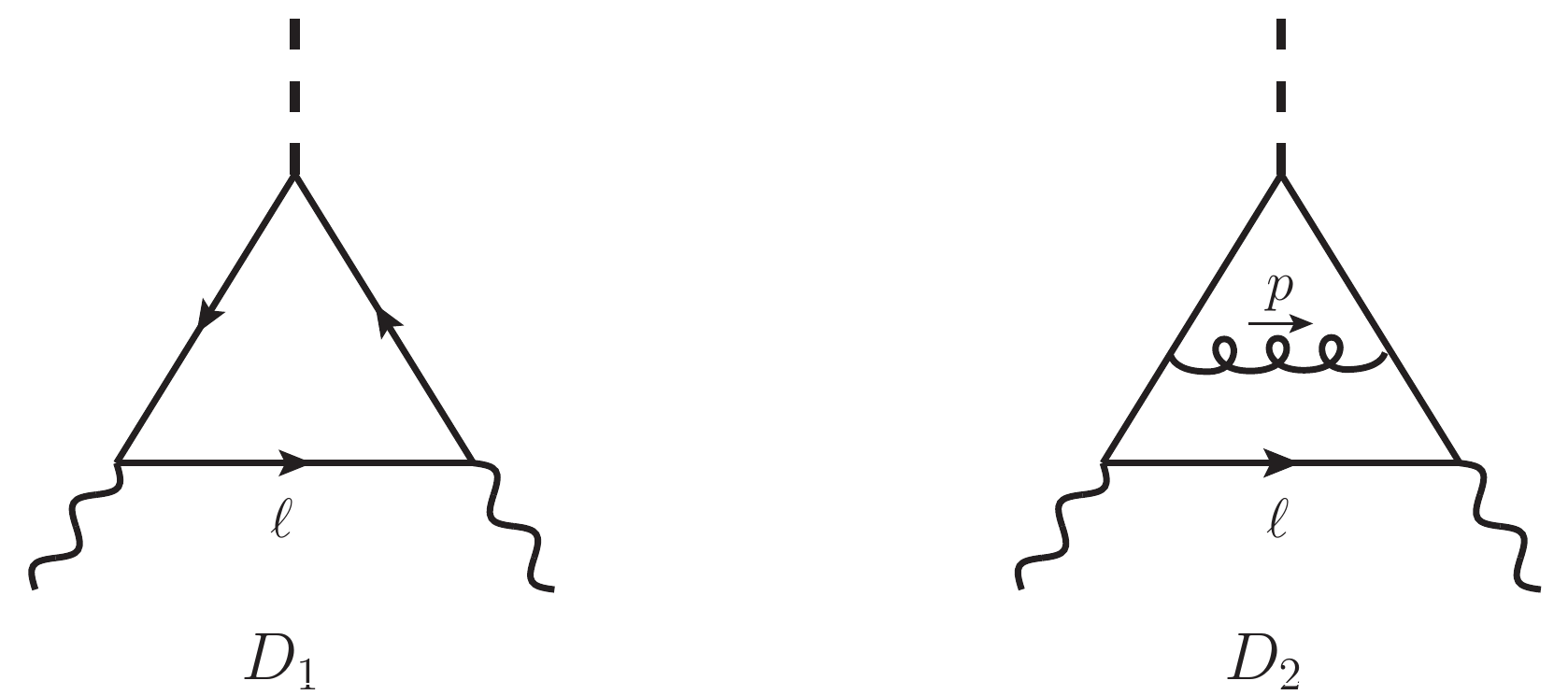} 
\vspace{2mm} 
\caption{\label{fig:new} 
Sample diagrams contributing to the $b$-quark induced $h\to\gamma\gamma$ amplitude.}
\end{center}
\end{figure}

To illustrate the region analysis, let us consider a one-loop graph and a sample two-loop diagram contributing to the $h\to\gamma\gamma$ decay amplitude in the full theory, see Figure~\ref{fig:new}. We find that the first graph receives leading-power contributions from exactly four regions, in which the loop momentum $\ell$ flowing between the two photon vertices is either hard ($h$), $n_1$-collinear ($c$), $n_2$-collinear ($\bar c$) or soft ($s$). Note that in the case where $\ell$ is soft the two remaining propagators carry hard-collinear momenta. The calculation of the individual contributions from the four regions requires rapidity regularization. Using the regularization scheme introduced in Section~\ref{sec:rapidity}, we find
\begin{equation}
\begin{aligned}
   D_1^{(h)} 
   &= {\cal M}_0 \left( \frac{1}{2\epsilon^2} - \frac{L_h}{2\epsilon} + \frac{L_h^2}{4}
    - 1 - \frac{\pi^2}{24} \right) , \\
   D_1^{(c)}  
   &= \frac{{\cal M}_0}{2} \left( \frac{1}{\epsilon} - L_m \right)
    \left( \frac{1}{\eta} + \ln\frac{-M_h^2-i0}{\nu^2} \right)
    = D_1^{(\bar c)} \,, \\
   D_1^{(s)} 
   &= {\cal M}_0 \left[ - \frac{1}{2\epsilon^2} + \frac{L_m}{2\epsilon} - \frac{L_m^2}{4} 
    + \frac{\pi^2}{24} - \left( \frac{1}{\epsilon} - L_m \right)
    \left( \frac{1}{\eta} + \frac12\ln\frac{-M_h^2-i0}{\nu^2} + \frac12\ln\frac{m_b^2}{\nu^2} \right) 
    \right] ,
\end{aligned}
\end{equation}
where $L_h=\ln(M_h^2/\mu^2)-i\pi$ and $L_m=\ln(m_b^2/\mu^2)$. The scale $\mu$ is associated with the dimensional regulator $\epsilon$, while $\nu$ is associated with the rapidity regulator $\eta$. Adding up the four contributions, we recover the correct result for the diagram,
\begin{equation}
   D_1 = {\cal M}_0 \left( \frac{L^2}{4} - 1 \right) ,
\end{equation}
with $L=L_h-L_m$, which is one half of the leading-order amplitude in (\ref{QCDresult}). For the two-loop diagram $D_2$, we find that leading-power contributions arise from the following regions of the loop momenta: $(\ell,p)\sim (h,h)$, $(c,h)$, $(c,c)$, $(\bar c,h)$, $(\bar c,\bar c)$, $(s,h)$, $(s,hc)$, $(s,\overline{hc})$, and $(s,s)$. In Section~\ref{subsec:5.1} we will calculate all contributions from the various regions at two-loop order and show that their sum correctly reproduces the QCD amplitude. This serves as a cross check that we have identified the leading momentum regions correctly.

\subsection{Matching onto SCET-1}
\label{subsec:2.1}

\begin{figure}[t]
\begin{center}
\includegraphics[width=0.5\textwidth]{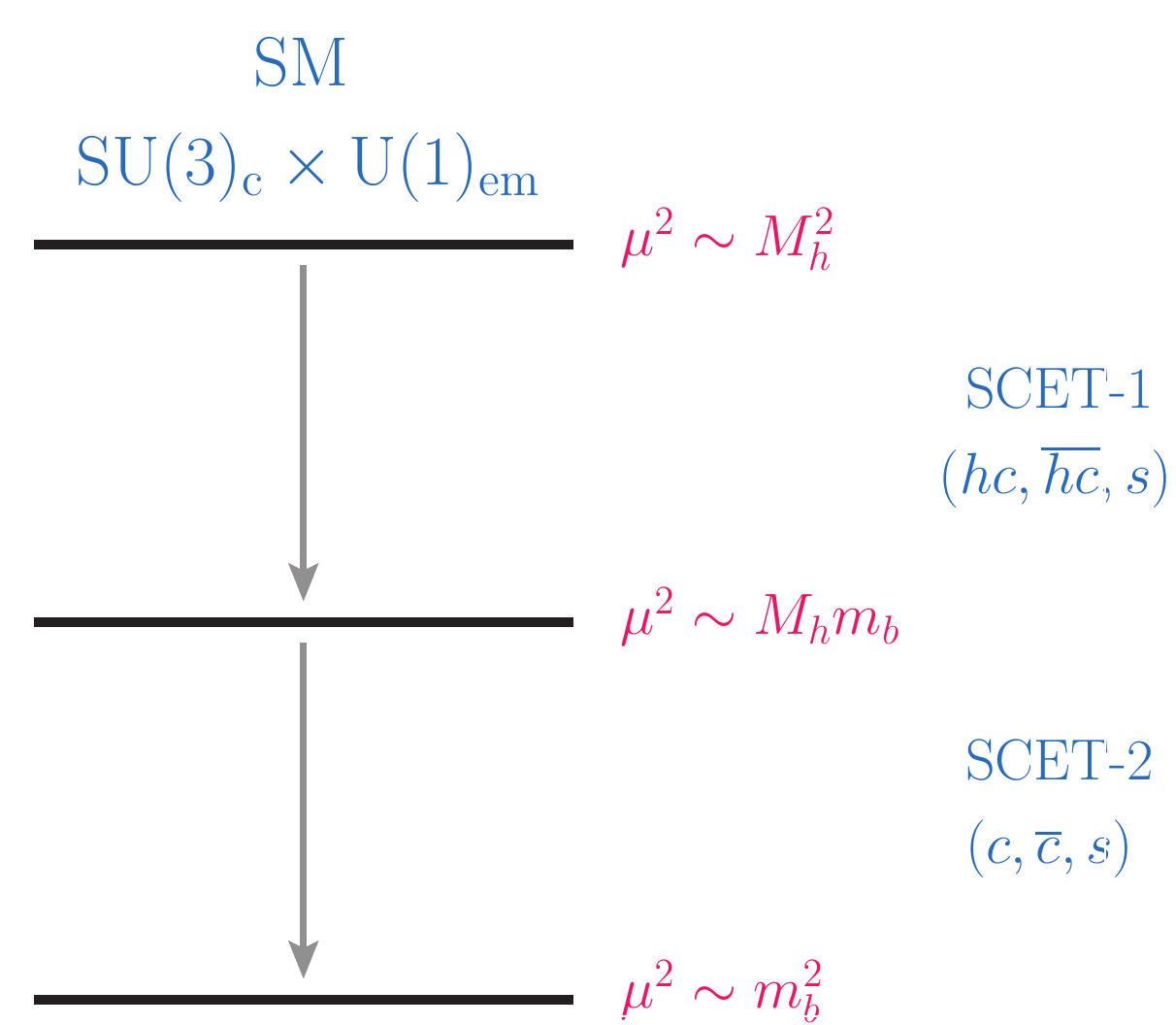} 
\vspace{2mm}
\caption{\label{fig:2step} 
Schematic illustration of the two-step matching procedure.}
\end{center}
\end{figure}

\begin{figure}[t]
\begin{center}
\includegraphics[width=0.45\textwidth]{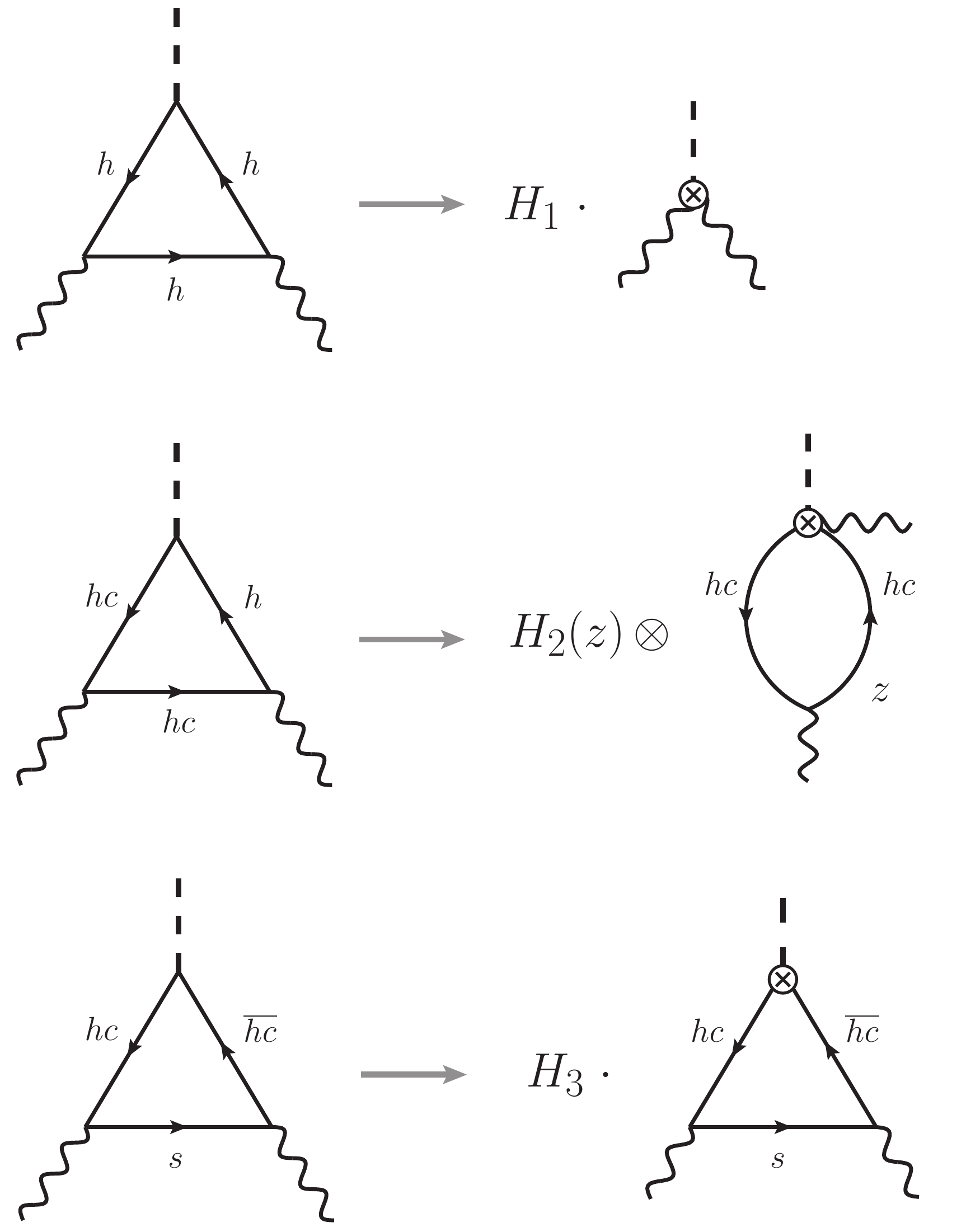} 
\vspace{2mm}
\caption{\label{fig:scet1} 
Leading regions of loop momenta contributing to the decay amplitude. The convolution symbol $\otimes$ in the second term means an integral over $z$.}
\end{center}
\end{figure}

One can describe the transition from QCD to SCET as a two-step process, as illustrated in Figure~\ref{fig:2step}. At a hard matching scale of order $\mu_h\sim M_h$, we match the SM (after electroweak symmetry breaking and with top-quarks integrated out) onto a version of SCET called SCET-1, which contains hard-collinear fields interacting with soft fields. The relevant momentum regions in this first effective theory are shown, at one-loop order, in Figure~\ref{fig:scet1}. From this it is straightforward to identify the associated SCET operators. The decay amplitude can be written in the form
\begin{equation}\label{fact0}
   {\cal M}_b(h\to\gamma\gamma)
   = \sum_i\,H_i\,\langle\gamma\gamma|\,O_i\,|h\rangle \,,
\end{equation}
where some of the products of hard matching coefficients and operator matrix elements are meant in the convolution sense. In the first diagram the loop momentum is hard, and so in the effective theory all propagators are shrunk to a point. The relevant SCET-1 operator $O_1$ contains the Higgs field coupled to two hard-collinear photon fields. In the second diagram the loop momentum connecting the two photons is $n_1$-hard-collinear, and hence the propagator between the second photon (with momentum $k_2$) and the Higgs boson is hard. In the effective theory this propagator is shrunk to a point. The relevant SCET operator $O_{2,n_1}$ contains the Higgs field, an $n_2$-hard-collinear photon and two $n_1$-hard-collinear quark fields. The latter two fields share the large momentum component $\bar n_1\cdot k_1$ of the photon, and hence the operator depends on a variable $z\in[0,1]$ denoting the momentum fraction carried by one of the two hard-collinear fields. Finally, in the third contribution the propagator between the two photons is soft and the two remaining propagators are hard-collinear. While soft gluon interactions with hard-collinear fields are contained in the leading-order SCET Lagrangian \cite{Bauer:2000yr,Bauer:2001yt}, interactions of soft quarks with hard-collinear quarks and gluons first arise at subleading power in SCET-1, more precisely at ${\cal O}(\lambda^{\frac12})$ \cite{Beneke:2002ph}. The relevant operator $O_3$ thus contains the time-ordered product of a scalar current made up of two hard-collinear quarks (coupled to the Higgs field) with two insertions of the subleading SCET Lagrangian coupling these quarks to soft quarks. 

The relevant SCET-1 operators needed to describe these contributions are 
\begin{equation}\label{Ondef}
\begin{aligned}
   O_1 &= \frac{m_b}{e_b^2}\,h(0)\,\A_{n_1}^{\perp\mu}(0)\,\A_{n_2,\mu}^\perp(0) \,, \\[-1.5mm]
   O_{2,n_1}(t) &= h(0)\,\bar\X_{n_1}(0)\,\gamma_\perp^\mu\,\frac{\nbsl_1}{2}\,\X_{n_1}(t\bar n_1)\,
    \A_{n_2,\mu}^\perp(0) \,, \\[0.5mm]
   O_3 &= T\,\Big\{ h(0)\,\bar\X_{n_1}(0)\,\X_{n_2}(0), i\!\int\!\!d^Dx\,{\cal L}_{q\,\xi_{n_1}}^{(1/2)}(x), 
    i\!\int\!\!d^Dy\,{\cal L}_{\xi_{n_2} q}^{(1/2)}(y) \Big\} + \mbox{h.c.} \,,
\end{aligned}
\end{equation}
and likewise for $O_{2,n_2}$ with $n_1\leftrightarrow n_2$. The factor $1/e_b^2$ in the definition of $O_1$ is chosen for later convenience. The Higgs field (a real scalar field after electroweak symmetry breaking) is denoted by $h$, whereas fields denoted by calligraphic letters are the so-called ``hard-collinear building blocks'' of SCET \cite{Bauer:2002nz,Hill:2002vw}. They are composite objects invariant under ``$n_i$-hard-collinear gauge transformations'', which preserve the scaling of the particle momenta shown in (\ref{hcfields}). These composite fields consist of hard-collinear fields dressed by hard-collinear Wilson lines
\begin{equation}
\begin{aligned}
   W_{n_i}^{(G)}(x) &= P\exp\left[ ig_s \int_{-\infty}^0\!ds\,\bar n_i\cdot G_{n_i}(x+s\bar n_i) \right] , \\
   W_{n_i}^{(A)}(x) &= P\exp\left[ ie_b \int_{-\infty}^0\!ds\,\bar n_i\cdot A_{n_i}(x+s\bar n_i) \right] 
\end{aligned}
\end{equation}
for QCD and QED interactions, respectively. The $n_i$-hard-collinear photon field is defined as \cite{Bauer:2002nz,Hill:2002vw}
\begin{equation}\label{calAdef}
   \A_{n_i}^\mu(x) = W_{n_i}^{(A)\dagger}(x) \big[ iD_{n_i}^\mu\,W_{n_i}^{(A)}(x) \big] 
   = e_b \int_{-\infty}^0\!ds\,\bar n_{i\nu} F_{n_i}^{\nu\mu}(x+s\bar n_i) \,,
\end{equation}
where $iD_{n_i}^\mu$ is a covariant derivative containing an $n_i$-hard-collinear gauge field. Note that the building block $\A_{n_i}^\mu$ includes a factor of the $b$-quark electric charge $e_b$ in its definition. The $n_i$-hard-collinear $b$-quark field is defined as
\begin{equation}
   \X_{n_i}(x) = \frac{\nsl_i\nbsl_i}{4}\,W_{n_i}^\dagger(x)\,b(x) \,, 
\end{equation}
where in general $W_{n_i}=W_{n_i}^{(G)} W_{n_i}^{(A)}$. Since we work to lowest order in QED interactions and the external photons have transverse polarization, we can however ignore the electromagnetic Wilson line in this definition. The advantage of using hard-collinear building blocks is that gauge invariance is explicit despite the fact that SCET is intrinsically non-local (through the appearance of the Wilson lines). Note the important fact that SCET fields living in different hard-collinear sectors (labeled by $n_1$ and $n_2$) can only interact with each other via the exchange of soft particles.

The operator $O_{2,n_1}(t)$ in (\ref{Ondef}) contains two hard-collinear quark fields in the same sector ($n_1$). When this appears, such fields do not need to live at the same spacetime point but can be delocalized along the $\bar n_1$ lightcone \cite{Beneke:2002ph}. The reason is that the derivative $(i\bar n_1\cdot\partial)={\cal O}(\lambda^0)$ is of leading order in SCET power counting and thus can be inserted an arbitrary number of times in such an operator. It is convenient to move from position to momentum space and define
\begin{equation}
   O_{2,n_1}(t) = \int\!d\omega\,e^{i\omega t}\,O_{2,n_1}(\omega) \,, \\
\end{equation}
where
\begin{equation}
   O_{2,n_1}(\omega) = h(0)\,\Big[ \bar\X_{n_1}(0)\,\gamma_\perp^\mu\,\frac{\nbsl_1}{2}\,
    \delta(\omega+i\bar n_1\cdot\partial)\,\X_{n_1}(0) \Big]\,\A_{n_2,\mu}^\perp(0) \,.
\end{equation}
The $\delta$-function enforces that the longitudinal momentum component $\bar n_1\cdot p$ of the outgoing hard-collinear anti-quark is equal to $\omega$. In SCET these large momentum components of hard-collinear particles are always positive and add up to the total hard-collinear momentum $\bar n_1\cdot{\cal P}_{n_1}$ in that sector, which in our case is the momentum $\bar n_1\cdot k_1$ of the external photon. We can thus parameterize $\omega=z_1\,\bar n_1\cdot k_1$ with $z_1\in[0,1]$. Note that the operator $O_{2,n_1}(\omega)$ is invariant under the rescaling transformation $\bar n_1^\mu\to\xi\,\bar n_1^\mu$, $n_1^\mu\to\xi^{-1} n_1^\mu$ with $\xi={\cal O}(1)$ \cite{Manohar:2002fd}, provided we also rescale $\omega\to\xi\,\omega$. It follows that the $h\to\gamma\gamma$ matrix element of $O_{2,n_1}(\omega)$ can only be a function of the invariant ratio $z_1=\omega/\bar n_1\cdot k_1$. Hence, the convolution
\begin{equation}
   \int_0^{\bar n_1\cdot k_1}\!d\omega\,H_{2,n_1}(\omega)\,\langle\gamma\gamma|\,O_{2,n_1}(\omega)\,|h\rangle
   \equiv \int_0^1\!dz_1\,H_2(z_1)\,\langle\gamma\gamma|\,O_2(z_1)\,|h\rangle
\end{equation}
can be expressed in terms of functions $H_2(z_1)$ and $\langle\gamma\gamma|\,O_2(z_1)\,|h\rangle$, whose only dependence on the reference vectors $n_1$ and $\bar n_1$ is through their argument $z_1=\omega/\bar n_1\cdot k_1$. The same functions appear in the convolution of $H_{2,n_2}(\omega)$ and $\langle\gamma\gamma|\,O_{2,n_2}(\omega)\,|h\rangle$ in the $n_2$-hard-collinear sector, but in this case they are evaluated at $z_2=\omega/\bar n_2\cdot k_2$.

As a side remark, we note that the photon matrix element of the $n_1$-hard-collinear fields contained in $O_{2,n_1}(t)$ is related to the leading-twist light-cone distribution amplitude $\phi_\gamma(u)$ of the photon \cite{Balitsky:1989ry,Ball:2002ps},
\begin{equation}
   \langle\gamma(k_1)|\,\bar\X_{n_1}(0)\,\gamma_\perp^\mu\,\frac{\nbsl_1}{2}\,\X_{n_1}(t\bar n_1)\,|0\rangle
   = - \frac{e_b}{2}\,\chi_m \langle\bar b b\rangle\,\bar n_1\cdot k_1\,\varepsilon_\perp^{\mu*}(k_1) 
    \int_0^1\!du\,e^{iut\bar n_1\cdot k_1}\,\phi_\gamma(u) \,.
\end{equation}
Here $\chi_m$ denotes the magnetic susceptibility and $\langle\bar b b\rangle$ is the $b$-quark vacuum condensate. It follows that $\langle\gamma\gamma|\,O_2(z)\,|h\rangle\propto\phi_\gamma(z)$. It can be shown rigorously that in the limit $\mu\to\infty$ the renormalized distribution amplitude approaches the asymptotic form 
\begin{equation}\label{phiasy}
   \lim_{\mu\to\infty} \phi_\gamma(z,\mu) = 6z(1-z) \,. 
\end{equation}
We will come back to the significance of this observation in Section~\ref{sec:7}.

The power-suppressed Lagrangian insertions in the definition of $O_3$ describe the coupling of a soft quark to hard-collinear fields at ${\cal O}(\lambda^{\frac12})$ in the SCET expansion parameter. In the notation of \cite{Beneke:2002ph}, one has
\begin{equation}
\begin{aligned}
   {\cal L}_{q\,\xi_{n_1}}^{(1/2)}(x)
   &= \bar q_s(x_-)\,W_{n_1}^\dagger(x)\,i\Dsl_{n_1}^\perp\,\xi_{n_1}(x) \,, \\
   {\cal L}_{\xi_{n_2} q}^{(1/2)}(y)
   &= \bar\xi_{n_2}(y) \big[i\Dsl_{n_2}^\perp\,W_{n_2}(y)\big]\,q_s(y_+) \,,
\end{aligned}
\end{equation}
where the soft quark fields $q_s$ must be multipole expanded for consistency \cite{Beneke:2002ph,Beneke:2002ni}, and we denote $x_-^\mu=(\bar n_1\cdot x)\,\frac{n_1^\mu}{2}$ and $y_+^\mu=(\bar n_2\cdot y)\,\frac{n_2^\mu}{2}$. Here $\xi_{n_1}$ is a collinear quark spinor subject to the constraint $\nsl_i\,\xi_{n_i}=0$, but not yet dressed by a Wilson line. When transformed into hard-collinear building blocks, the corresponding expressions are
\begin{equation}
\begin{aligned}
   {\cal L}_{q\,\xi_{n_1}}^{(1/2)}(x)
   &= \bar q_s(x_-) \big[ \Asl_{n_1}^\perp(x) + \Gsl_{n_1}^\perp(x) \big]\,\X_{n_1}(x) \,, \\
   {\cal L}_{\xi_{n_2} q}^{(1/2)}(y)
   &= \bar\X_{n_2}(y) \big[ \Asl_{n_2}^\perp(y) + \Gsl_{n_2}^\perp(y) \big]\,q_s(y_+) \,,
\end{aligned}
\end{equation}
where $\A_{n_1}^\mu$ and $\G_{n_1}^\mu$ are the building blocks for the hard-collinear photon and gluon fields, respectively. $\G_{n_i}^\mu$ is defined in analogy with the first relation for $\A_{n_i}^\mu$ in (\ref{calAdef}). Note that in the subleading-power Lagrangians the different hard-collinear fields appear at the same spacetime point, unlike the more general case in (\ref{Ondef}). Also, it can be shown that the subleading-power Lagrangians are not renormalized \cite{Beneke:2002ph}.

Let us now analyze the power counting of the three operators in (\ref{Ondef}). With our power counting, the hard-collinear quark, (transverse) photon and gluon fields all scale like $\lambda^{\frac12}$, whereas the soft quark field scales like $\lambda^{\frac32}$. Using that $m_b\sim\lambda$, it follows that $O_1\sim\lambda^2$ and $O_3\sim\lambda^2$, whereas it would appear that $O_{2,n_i}\sim\lambda^{\frac32}$. However, the $h\to\gamma\gamma$ matrix element of $O_{2,n_i}$ requires a mass insertion of $m_b$ in order to be non-zero, and an ${\cal O}(\lambda)$ quark mass term in the hard-collinear Lagrangian corresponds to a power-suppressed interaction of ${\cal O}(\lambda^{\frac12})$. We refrain from writing this interaction in the form of a time-ordered product, since after matching to SCET-2 the quark mass term will be promoted to a leading-order interaction. As a consequence, it follows that $O_{2,n_i}\sim\lambda^2$ has the same scaling as the other operators. 

Let us note, for comparison, that the coupling of a heavy vector boson $Z'$ to a pair of hard-collinear quarks would be described by the SCET-1 operator $Z_\mu'\,\bar\X_{n_1}\X_{n_2}\sim\lambda$. Hence, it follows that the three operators in (\ref{Ondef}) are of subleading order in SCET power counting. It is only for this reason that we obtain a factorization theorem involving a sum of three types of operators, as indicated in (\ref{fact0}).

At leading order in the SCET expansion, the hard-collinear quark and gluon fields in the operators in (\ref{Ondef}) can interact with soft gluons (and photons) through eikonal couplings involving the components $n_i\cdot G_s$ (and $n_i\cdot A_s$) of the soft gauge fields. These couplings can be removed from the Lagrangian by means of the field redefinitions \cite{Bauer:2001yt}
\begin{equation}\label{decoupling}
\begin{aligned}
   \X_{n_1}(x) &\to S_{n_1}(x_-)\,\X_{n_1}(x) \,, \\
   \G_{n_1}^{\perp\mu}(x) &\to S_{n_1}^{(G)}(x_-)\,\G_{n_1}^{\perp\mu}(x)\,S_{n_1}^{(G)\dagger}(x_-) \,,
\end{aligned}
\end{equation}
and similarly for the $n_2$-hard-collinear fields with $x_-$ replaced by $x_+$. Here $S_{n_i}$ are soft Wilson lines defined as
\begin{equation}
\begin{aligned}
   S_{n_i}^{(G)}(x) &= P\exp\left[ ig_s \int_{-\infty}^0\!dt\,n_i\cdot G_s(x+t n_i) \right] , \\
   S_{n_i}^{(A)}(x) &= P\exp\left[ ie_b \int_{-\infty}^0\!dt\,n_i\cdot A_s(x+t n_i) \right] ,
\end{aligned}
\end{equation}
and $S_{n_1}=S_{n_i}^{(G)} S_{n_i}^{(A)}$. After the decoupling transformations (\ref{decoupling}) the fields in each sector ($n_1$-hard-collinear, $n_2$-hard-collinear and soft) only interact among themselves, but there are no interactions connecting two sectors. The operators $O_1$ and $O_{2,n_i}$ are left invariant under the transformations in (\ref{decoupling}), whereas the operator $O_3$ can be factorized in the form 
\begin{equation}\label{O3val}
\begin{aligned}
   O_3 &= h(0) \int\!\!d^Dx\!\int\!\!d^Dy\,
    T\,\Big\{ \big[\big(\Asl_{n_1}^\perp(x)+\Gsl_{n_1}^\perp(x)\big)\,\X_{n_1}(x)\big]^{\alpha i}\,
    \bar\X_{n_1}^{\beta j}(0) \Big\} \\
   &\quad\times T\,\Big\{ \X_{n_2}^{\beta k}(0) 
    \big[\bar\X_{n_2}(y)\,\big(\Asl_{n_2}^\perp(y)+\Gsl_{n_2}^\perp(y)\big)\big]^{\gamma l} \Big\} \\[2mm]
   &\quad\times T\,\Big\{ [S_{n_2}^\dagger(y_+)\,q_s(y_+)]^{\gamma l}\,[\bar q_s(x_-)\,S_{n_1}(x_-)]^{\alpha i}\,
    [S_{n_1}^\dagger(0)\,S_{n_2}(0)]^{jk} \Big\} + \mbox{h.c.} \,, 
\end{aligned}
\end{equation}
where we have written out the color ($i,j,k,l$) and spinor ($\alpha,\beta,\gamma$) indices explicitly and taken into account a factor $(-1)$ arising from the anti-commutators of fermionic fields. 

\subsection{Matching onto SCET-2}

Eventually, we are interested in evaluating the $h\to\gamma\gamma$ matrix element in (\ref{fact0}) using on-shell photon states. The various operators and Wilson coefficients in (\ref{fact0}) can be evolved to scales below the hard matching scale $\mu_h\sim M_h$ using renormalization-group (RG) evolution equations in SCET-1. At a hard-collinear matching scale of order $\mu_{hc}\sim\sqrt{M_h m_b}$, we match the theory constructed in the previous subsection onto another version of SCET called SCET-2. It contains collinear fields and soft fields, which do not interact with each other. For the operators $O_1$ and $O_{2,n_i}$ this matching is trivial in the sense that we simply need to replace the various hard-collinear fields with collinear fields, which are defined in a completely analogous way. Note that, as mentioned above, the quark mass now appears in the leading-order SCET Lagrangian. The collinear quark and transverse gauge fields all scale like $\lambda$ in SCET-2, and hence the operators in (\ref{fact0}) now scale like $\lambda^3$. 

\begin{figure}[t]
\begin{center}
\includegraphics[width=0.55\textwidth]{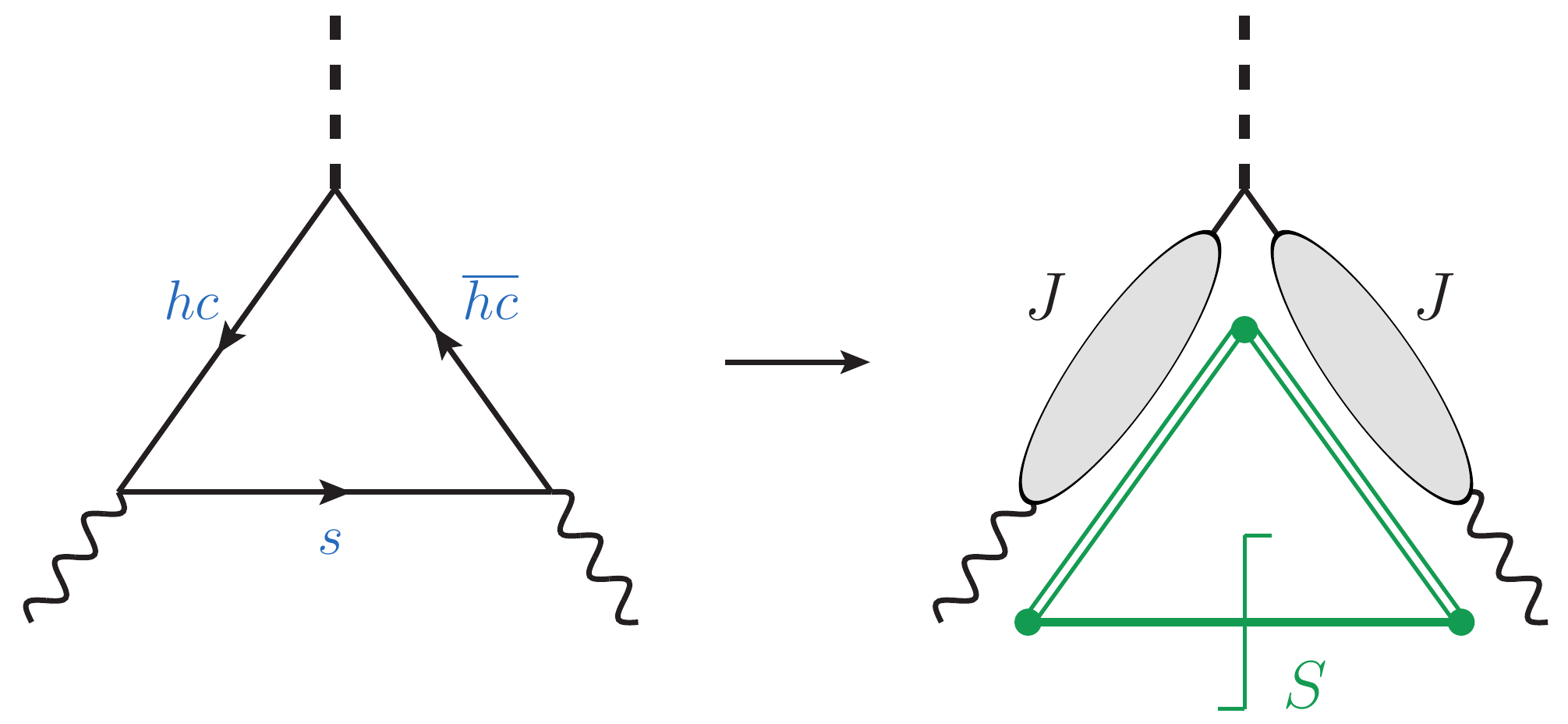}
\vspace{2mm}
\caption{\label{fig:regions} 
Factorization of the matrix element $\langle O_3\rangle$ in SCET-2.}
\end{center}
\end{figure}

The matching relation for the operator $O_3$ is non-trivial, because the coupling of a soft quark to a collinear photon produces hard-collinear quanta, which must be integrated out at the hard-collinear scale. This is illustrated in Figure~\ref{fig:regions}. In this process two jet functions arise, one in each hard-collinear sector. There is a crucial difference here with regard to the case of electroweak Sudakov resummation at leading power, as studied in \cite{Chiu:2007yn,Chiu:2007dg}. The exchange of soft gluons between two hard-collinear quarks is entirely described in terms of emissions from soft Wilson lines, without the appearance of non-trivial matching coefficients. In the present case, however, non-trivial matching coefficients appear (see also \cite{Moult:2019mog,Moult:2019uhz}). Integrating out the hard-collinear fields gives rise to a generalized jet function $J$, which we define as
\begin{equation}\label{jet1}
\begin{aligned}
   &\langle\gamma(k_1)|\,T\,\big[\big(\Asl_{n_1}^\perp(x)+\Gsl_{n_1}^\perp(x)\big)\,\X_{n_1}(x)\big]^{\alpha i}\,
    \bar\X_{n_1}^{\beta j}(r)\,|0\rangle \\
   &= e_b\,\delta^{ij}\,\Big[\rlap/\varepsilon_\perp^*(k_1)\,\frac{\nsl_1}{2} \Big]^{\alpha\beta} 
    \int\frac{d^Dp}{(2\pi)^D}\,\frac{i\bar n_1\cdot p}{p^2+i0}\,J\big(p^2,(p-k_1)^2\big)\,
    e^{-ip\cdot(x-r)+i k_1\cdot x} \,, 
\end{aligned}
\end{equation}
such that $J=1+{\cal O}(\alpha_s)$. Here $p$ denotes the momentum leaving the Higgs vertex at position $r$, and $(p-k_1)$ is the momentum flowing into the vertex at point $x$. The function $J$ can depend on the two invariant scalar products $p^2$ and $(p-k_1)^2$, where $k_1$ is the momentum of the external photon, and thus $k_1^2=0$. Note that the above definition refers to the massless theory, because the quark mass must be set to zero for the hard-collinear fields. 

The corresponding matrix element of the hard-collinear fields in the second line of (\ref{O3val}) can be obtained from the expression above by hermitian conjugation, exchange of $n_1\leftrightarrow n_2$ and $k_1\leftrightarrow k_2$, and using crossing symmetry (i.e.\ moving the photon from the initial to the final state). This yields 
\begin{equation}\label{jet2}
\begin{aligned}
   &\langle\gamma(k_2)|\,T\,\X_{n_2}^{\beta k}(r) 
    \big[\bar\X_{n_2}(y)\,\big(\Asl_{n_2}^\perp(y)+\Gsl_{n_2}^\perp(y)\big)\big]^{\gamma l}\,|0\rangle \\
   &= e_b\,\delta^{kl}\,\Big[\,\frac{\nsl_2}{2}\,\rlap/\varepsilon_\perp^*(k_2) \Big]^{\beta\gamma}
    \int\frac{d^Dp}{(2\pi)^D}\,\frac{i\bar n_2\cdot p}{p^2+i0}\,J\big(p^2,(p+k_2)^2\big)\,
    e^{-ip\cdot(r-y)+i k_2\cdot y} \,. 
\end{aligned}
\end{equation}

We now insert the expressions (\ref{jet1}) and (\ref{jet2}) into the matrix element (\ref{O3val}) and perform the relevant contractions of color and spinor indices. We are then led to define the soft matrix element in the form 
\begin{equation}
\begin{aligned}
   &\frac{e_b^2}{\pi}\,\langle 0|\,T\,\mbox{Tr}\,S_{n_2}(0,y_+)\,q_s^\gamma(y_+)\,
    \bar q_s^\alpha(x_-)\,S_{n_1}(x_-,0)\,|0\rangle \\
   &= i\int\frac{d^D\ell}{(2\pi)^D}\,e^{-i\ell\cdot(y_+ - x_-)}\,
    \bigg[ {\cal S}_1(\ell) + \rlap/\ell\,{\cal S}_2(\ell)
    + \frac{\nsl_1}{n_1\cdot\ell}\,{\cal S}_3(\ell) + \frac{\nsl_2}{n_2\cdot\ell}\,{\cal S}_4(\ell) \\
   &\hspace{4.64cm}\mbox{}+ \frac{\rlap/\ell\nsl_1}{n_1\cdot\ell}\,{\cal S}_5(\ell)
    + \frac{\nsl_2\rlap/\ell}{n_2\cdot\ell}\,{\cal S}_6(\ell) 
    + \frac{\nsl_2\nsl_1}{4}\,{\cal S}_7(\ell) 
    + \frac{\nsl_2\rlap/\ell\nsl_1}{2}\,{\cal S}_8(\ell) \bigg]^{\gamma\alpha} \,,
\end{aligned}
\end{equation}
where the trace in the first line is over color indices, and we have introduced the {\em finite-length\/} soft Wilson lines
\begin{equation}\label{eq25}
\begin{aligned}
   S_{n_1}^{(G)}(x_-,0) \equiv S_{n_1}^{(G)}(x_-)\,S_{n_1}^{(G)\dagger}(0)
   &= P\exp\left[ ig_s \int_0^{\bar n_1\cdot x/2}\!dt\,n_1\cdot G_s(t n_1) \right] , \\
   S_{n_2}^{(G)}(0,y_+) \equiv S_{n_2}^{(G)}(0)\,S_{n_2}^{(G)\dagger}(y_+)
   &= P\exp\left[ ig_s \int_{\bar n_2\cdot y/2}^0\!dt\,n_2\cdot G_s(t n_2) \right] , 
\end{aligned}
\end{equation}
and similarly for the electromagnetic Wilson lines. Here the path ordering is such that the fields appear in the same order as the arguments in the functions. For example, in the expression for $S_{n_1}(x_-,0)$ the fields are ordered according to $t$-values, such that fields at $t=\bar n_1\cdot x/2$ appear to the left and fields at $t=0$ appear to the right. Reparameterization invariance \cite{Manohar:2002fd} implies that the structure functions ${\cal S}_i(\ell)$ can only depend on the invariants $n_1\cdot\ell\,n_2\cdot\ell$ and $\ell_\perp^2$. The prefactor $e_b^2/\pi$ is included for later convenience. The multipole expansion ensures that only those components of the soft momentum which are commensurate with the relevant hard-collinear momenta enter at the vertices $x$ and $y$ \cite{Beneke:2002ph,Beneke:2002ni}. Concretely, the momentum $(n_1\cdot\ell)\,\frac{\bar n_1^\mu}{2}$ leaves the vertex at $x$, where the soft quark connects to $n_1$-hard-collinear fields, while $(n_2\cdot\ell)\,\frac{\bar n_2^\mu}{2}$ flows into the vertex at $y$, where the soft quark connects to $n_2$-hard-collinear fields. 

When the soft and jet functions are combined according to (\ref{O3val}), one finds that all structures except for ${\cal S}_1$ vanish. Hence, after matching onto SCET-2 the $h\to\gamma\gamma$ matrix element of $O_3$ assumes the factorized form (omitting the photon polarization vectors for simplicity)
\begin{equation}
\begin{aligned}
   \langle\gamma\gamma|\,O_3\,|h\rangle 
   &= 2\pi i\,\bar n_1\cdot k_1\,\bar n_2\cdot k_2\,g_\perp^{\mu\nu} \int\frac{d^D\ell}{(2\pi)^D}\,
    \frac{J\big(\bar n_1\cdot k_1\,n_1\cdot\ell,0\big)}{\bar n_1\cdot k_1\,n_1\cdot\ell+i0}\,
    \frac{J\big(-\bar n_2\cdot k_2\,n_2\cdot\ell,0\big)}{-\bar n_2\cdot k_2\,n_2\cdot\ell+i0}\,{\cal S}_1(\ell) \\
   &\quad\mbox{}+ (1\leftrightarrow 2) \,. 
\end{aligned}
\end{equation}
The matrix element of the hermitian conjugate operator in (\ref{O3val}) is obtained by interchanging the subscripts 1 and 2 everywhere. The two matrix elements are related to each other by changing the sign of the loop momentum $\ell$, which leaves the soft function ${\cal S}_1(\ell)$ unchanged. Hence, the term $(1\leftrightarrow 2)$ simply leads to a factor~2. Note that the second argument of the function $J$ is zero in both cases, since the relevant momentum is equal to the multipole expanded soft momentum, e.g.\ $(n_1\cdot\ell)\,\frac{\bar n_1^\mu}{2}$ for the case of the first jet function, which is light-like. We will from now on only display the first argument for brevity. Finally, because of the multipole expansion, the transverse components of the loop momentum only enter in the soft function. We can thus rewrite the answer in the form (with $\ell_+\equiv n_1\cdot\ell$ and $\ell_-\equiv n_2\cdot\ell$)
\begin{equation}\label{Ondef2}
   \langle\gamma\gamma|\,O_3\,|h\rangle 
   = \frac{ig_\perp^{\mu\nu}}{2\pi} \int_{-\infty}^\infty\!d\ell_+\!\int_{-\infty}^\infty\!d\ell_-\,
    \frac{J(\bar n_1\cdot k_1\,\ell_+)}{\ell_+ +i0}\,
    \frac{J(-\bar n_2\cdot k_2\,\ell_-)}{-\ell_- +i0}\,{\cal S}(\ell_+\ell_-) \,,
\end{equation}
where
\begin{equation}
   {\cal S}(\ell_+\ell_-) \equiv \int\frac{d^{D-2}\ell_\perp}{(2\pi)^{D-2}}\,{\cal S}_1(\ell) \,. 
\end{equation}
This result clearly shows that the matrix element depends on both the soft and the hard-collinear scale, in accordance with Figure~\ref{fig:regions}. 

\begin{figure}[t]
\begin{center}
\includegraphics[width=0.65\textwidth]{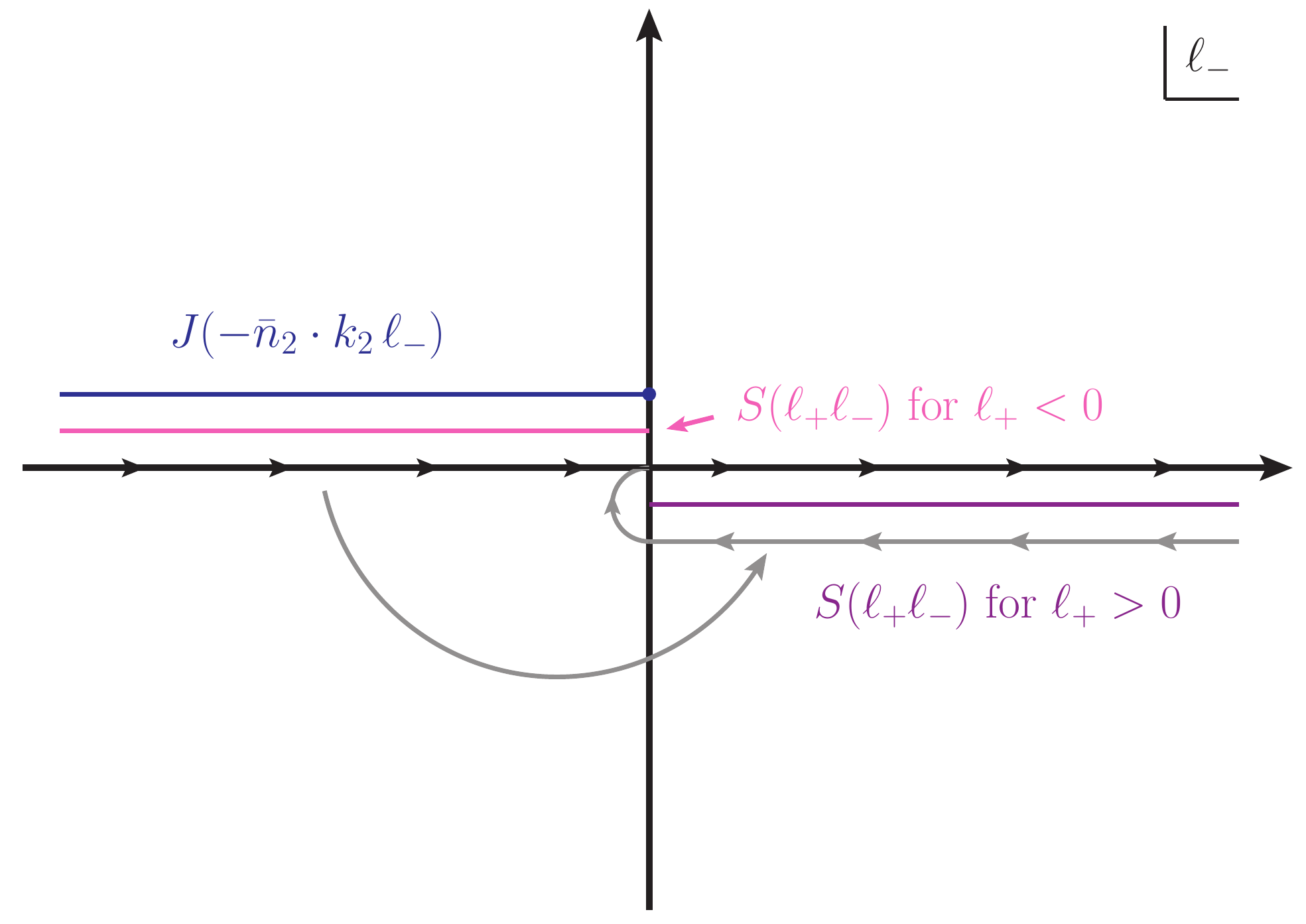}
\vspace{2mm}
\caption{\label{fig:poles} 
Singularities of the integral (\ref{Ondef2}) in the complex $\ell_-$ plane.}
\end{center}
\end{figure}

In Figure~\ref{fig:poles} we show the analytic structure of the integrand in (\ref{Ondef2}) in the complex $\ell_-$ plane. The jet function $J(-\bar n_2\cdot k_2\,\ell_-)/(-\ell_- +i0)$ has a pole at $\ell_-=i0$ and a cut along (and slightly above) the negative real axis starting at the origin. The integrated soft function has a cut starting at $\ell_+\ell_-=-i0$ and extending to infinity. For $\ell_+<0$, this cut is also located above the real axis. The contour can then be closed in the lower half-plane (the dimensional regulator ensures that the contribution from the half-circle at infinity vanishes) and hence the integral yields zero. For $\ell_+>0$, however, the discontinuities of the soft function lie in the lower half-plane and thus the integral is non-zero. It follows that we can express the matrix element in the form 
\begin{equation}\label{Ondef3a}
   \langle\gamma\gamma|\,O_3\,|h\rangle 
   = g_\perp^{\mu\nu} \int_0^\infty\!\frac{d\ell_+}{\ell_+}\!\int_0^\infty\!\frac{d\ell_-}{\ell_-}\,
    J(\bar n_1\cdot k_1\,\ell_+)\,J(-\bar n_2\cdot k_2\,\ell_-)\,S(\ell_+\ell_-) \,,
\end{equation}
where
\begin{equation}\label{SDisc}
   S(\ell_+\ell_-) 
   = \frac{1}{2\pi i}\,\Big[ {\cal S}(\ell_+\ell_- +i0) - {\cal S}(\ell_+\ell_- -i0) \Big]
\end{equation}
is the discontinuity of the soft function ${\cal S}(\ell_+\ell_-)$. An analogous discussion can be made in the complex $\ell_+$ plane. In this case one obtains (\ref{Ondef3a}) with $\bar n_1\cdot k_1\leftrightarrow\bar n_2\cdot k_2$ and $\ell_+\leftrightarrow\ell_-$ interchanged, i.e.\ with different signs in the arguments of the jet functions. It follows that both expressions are equivalent. However, we will see later that the matrix element of $O_3$ is ill-defined without an additional rapidity regulator, and this regulator would break the exchange symmetry if we would use either (\ref{Ondef3a}) or the corresponding other result. The symmetry can be restored by using the symmetric form
\begin{equation}\label{Ondef3}
\begin{aligned}
   \langle\gamma\gamma|\,O_3\,|h\rangle 
   &= \frac{g_\perp^{\mu\nu}}{2} \int_0^\infty\!\frac{d\ell_+}{\ell_+}\!\int_0^\infty\!\frac{d\ell_-}{\ell_-}\,
    \Big[ J(\bar n_1\cdot k_1\,\ell_+)\,J(-\bar n_2\cdot k_2\,\ell_-) \\
   &\hspace{4.35cm}\mbox{}+ J(-\bar n_1\cdot k_1\,\ell_+)\,J(\bar n_2\cdot k_2\,\ell_-) \Big]\,
    S(\ell_+\ell_-) \,.
\end{aligned}
\end{equation}

At this stage, the factorization theorem for the $b$-quark induced contribution to the $h\to\gamma\gamma$ decay amplitude can be written in the form
\begin{equation}\label{fact1}
   {\cal M}_b(h\to\gamma\gamma)
   = H_1\,\langle\gamma\gamma|\,O_1\,|h\rangle 
    + \left[ \int_0^1\!dz_1\,H_2(z_1)\,\langle\gamma\gamma|\,O_2(z_1)\,|h\rangle + (z_1\to z_2) \right]
    + H_3\,\langle\gamma\gamma|\,O_3\,|h\rangle \,,
\end{equation}
with $\langle\gamma\gamma|\,O_3\,|h\rangle$ given in (\ref{Ondef3}). Subtleties arise from endpoint-divergent convolution integrals in all but the first term, which need to be properly identified and regularized. To see how the problem arises, we now present explicit expressions for the Wilson coefficients and operator matrix elements at NLO in $\alpha_s$.

\section{Hard matching coefficients}
\label{sec:3}

\begin{figure}[t]
\begin{center}
\includegraphics[width=\textwidth]{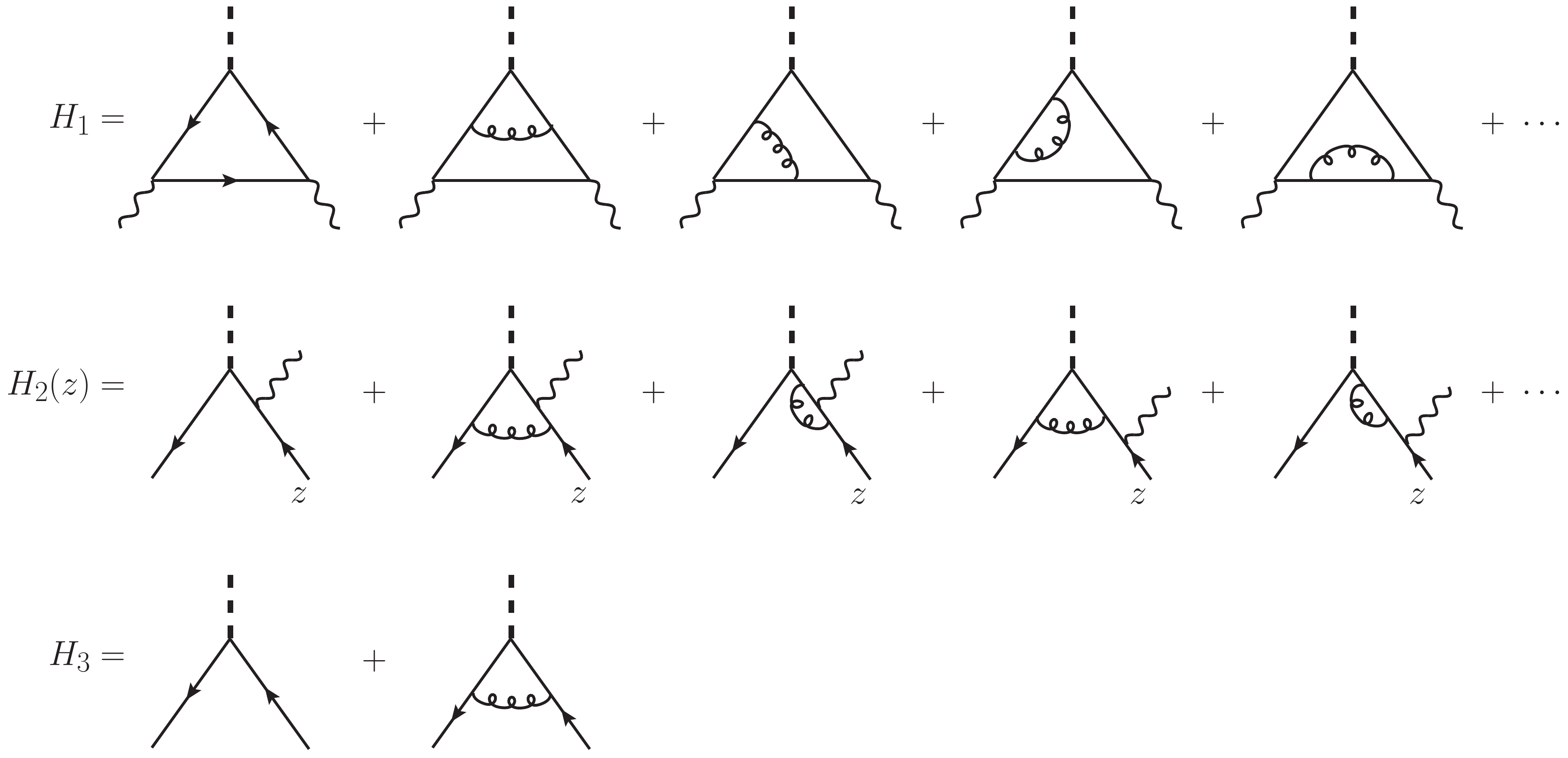}
\vspace{2mm}
\caption{\label{fig:hard} 
Representative Feynman diagrams contributing to the hard matching coefficients $H_i$ up to ${\cal O}(\alpha_s)$. In the case of $H_1$ we do not show the mirror graphs of the third and fourth diagram, in which the gluon is attached on the other side. In the case of $H_2$ we omit the corresponding diagrams with the photon attached to the other quark line.}
\end{center}
\end{figure}

We begin by describing the calculation of the Wilson coefficients $H_i$, which contain the hard matching contributions obtained when matching QCD to SCET. The relevant Feynman diagrams up to one-loop order are shown in Figure~\ref{fig:hard}. Evaluating these diagrams in dimensional regularization with $D=4-2\epsilon$ spacetime dimensions and in the $\overline{\rm MS}$ scheme, we obtain 
\begin{equation}\label{Hires}
\begin{aligned}
   H_1 &= \frac{y_{b,0}}{\sqrt2}\,\frac{N_c\alpha_{b,0}}{\pi} \left( -M_h^2 -i0 \right)^{-\epsilon}
    e^{\epsilon\gamma_E}\,(1-3\epsilon)\,\frac{2\Gamma(1+\epsilon)\,\Gamma^2(-\epsilon)}{\Gamma(3-2\epsilon)} \\
   &\quad\times \bigg\{ 1 - \frac{C_F\alpha_{s,0}}{4\pi} \left( -M_h^2 -i0 \right)^{-\epsilon}
    e^{\epsilon\gamma_E}\,\frac{\Gamma(1+2\epsilon)\,\Gamma^2(-2\epsilon)}{\Gamma(2-3\epsilon)} \\
   &\hspace{1.45cm}\times \bigg[ \frac{2(1-\epsilon)(3-12\epsilon+9\epsilon^2-2\epsilon^3)}{1-3\epsilon} 
    + \frac{8}{1-2\epsilon}\,\frac{\Gamma(1+\epsilon)\,\Gamma^2(2-\epsilon)\,\Gamma(2-3\epsilon)}%
                                  {\Gamma(1+2\epsilon)\,\Gamma^3(1-2\epsilon)} \\
   &\hspace{2.1cm}\mbox{}- \frac{4(3-18\epsilon+28\epsilon^2-10\epsilon^3-4\epsilon^4)}{1-3\epsilon}\,
     \frac{\Gamma(2-\epsilon)}{\Gamma(1+\epsilon)\,\Gamma(2-2\epsilon)} \bigg] \bigg\} \,, \\
   H_2(z) &= \frac{y_{b,0}}{\sqrt2}\,\bigg\{ \frac{1}{z} 
    + \frac{C_F\alpha_{s,0}}{4\pi} \left( -M_h^2 -i0\right)^{-\epsilon}
    e^{\epsilon\gamma_E}\,\frac{\Gamma(1+\epsilon)\,\Gamma^2(-\epsilon)}{\Gamma(2-2\epsilon)} \\
   &\hspace{1.9cm}\times
    \left[ \frac{2-4\epsilon-\epsilon^2}{z^{1+\epsilon}} - \frac{2(1-\epsilon)^2}{z}
     -2(1-2\epsilon-\epsilon^2)\,\frac{1-z^{-\epsilon}}{1-z} \right] \bigg\} + (z\to 1-z) \,, \\
   H_3 &= - \frac{y_{b,0}}{\sqrt 2} \left[ 1 
    - \frac{C_F\alpha_{s,0}}{4\pi} \left( -M_h^2 -i0\right)^{-\epsilon}
    e^{\epsilon\gamma_E}\,2(1-\epsilon)^2\,
    \frac{\Gamma(1+\epsilon)\,\Gamma^2(-\epsilon)}{\Gamma(2-2\epsilon)} \right] .
\end{aligned}
\end{equation}
Here $y_{b,0}$ is the bare $b$-quark Yukawa coupling, and $\alpha_{s,0}$ and $\alpha_{b,0}$ denote the bare QCD and electromagnetic couplings, respectively. These expressions are exact to all orders in $\epsilon$. Note that the factor $\alpha_{b,0}$ included in $H_1$ arises because we have defined $O_1$ without including any electric charges in the operator. When the bare couplings are renormalized according to
\begin{equation}\label{alpharen}
   \alpha_{b,0} = \mu^{2\epsilon} \alpha_b(\mu) + {\cal O}(\alpha^2) \,, \qquad
   \alpha_{s,0} = \mu^{2\epsilon} \alpha_s(\mu) + {\cal O}(\alpha_s^2) \,,
\end{equation}
this ensures that the three Wilson coefficients depend on the dimensionless ratio $(-M_h^2-i0)/\mu^2$.

We will see in the next section that the matrix element $\langle\gamma\gamma|\,O_2(z)\,|h\rangle$ is symmetric under the exchange $z\leftrightarrow(1-z)$, as is the case for $H_2(z)$. We can therefore rewrite 
\begin{equation}\label{rewrite}
   \int_0^1\!dz\,H_2(z)\,\langle\gamma\gamma|\,O_2(z)\,|h\rangle
   = 2\int_0^1\!\frac{dz}{z}\,\bar H_2(z)\,\langle\gamma\gamma|\,O_2(z)\,|h\rangle \,,
\end{equation}
with
\begin{equation}\label{H2bar}
\begin{aligned}
   \bar H_2(z) &= \frac{y_{b,0}}{\sqrt2}\,\bigg\{ 1 + \frac{C_F\alpha_{s,0}}{4\pi} \left( - M_h^2 \right)^{-\epsilon}
    e^{\epsilon\gamma_E}\,\frac{\Gamma(1+\epsilon)\,\Gamma^2(-\epsilon)}{\Gamma(2-2\epsilon)} \\
   &\hspace{1.8cm}\times
    \Big[ (2-4\epsilon-\epsilon^2)\,z^{-\epsilon} - 2(1-\epsilon)^2
     -2(1-2\epsilon-\epsilon^2)\,\big[ 1-(1-z)^{-\epsilon} \big] \Big] \bigg\} \,.
\end{aligned}
\end{equation}
We will also see that the matrix element is $z$-independent at leading order. It is then obvious that the integrals over $z_i$ in (\ref{fact1}) diverge at the endpoints $z_i=0$ and $z_i=1$. With the help of (\ref{rewrite}), the divergence is concentrated at $z_i=0$. Moreover, we will find later that the integrals over $\ell_\pm$ in (\ref{Ondef3}) diverge when $\ell_+\to\infty$ or $\ell_-\to\infty$ at fixed $\ell_+\ell_-$. These divergences are not regularized by the dimensional regulator $\epsilon$ and require additional (rapidity) regulators. This will be discussed in more detail in Section~\ref{sec:rapidity}.

\section{Operator matrix elements}
\label{sec:4}

In the next step, we need the matrix elements of the SCET operators in (\ref{Ondef}). We have calculated these matrix elements at NLO in $\alpha_s$ in both SCET-1 and SCET-2. In SCET-1, the hard-collinear scaling can be enforced by taking the external photons off-shell, with virtualities $(-k_i^2)={\cal O}(M_h m_b)$. The corresponding expressions for the matrix elements are presented in Appendix~\ref{app:A}. Using these results, it should be possible to derive the RG evolution equations for the hard matching coefficients in the region between the hard scale $M_h$ and the hard-collinear scale $\sqrt{M_h m_b}$.

\begin{figure}[t]
\begin{center}
\includegraphics[width=0.7\textwidth]{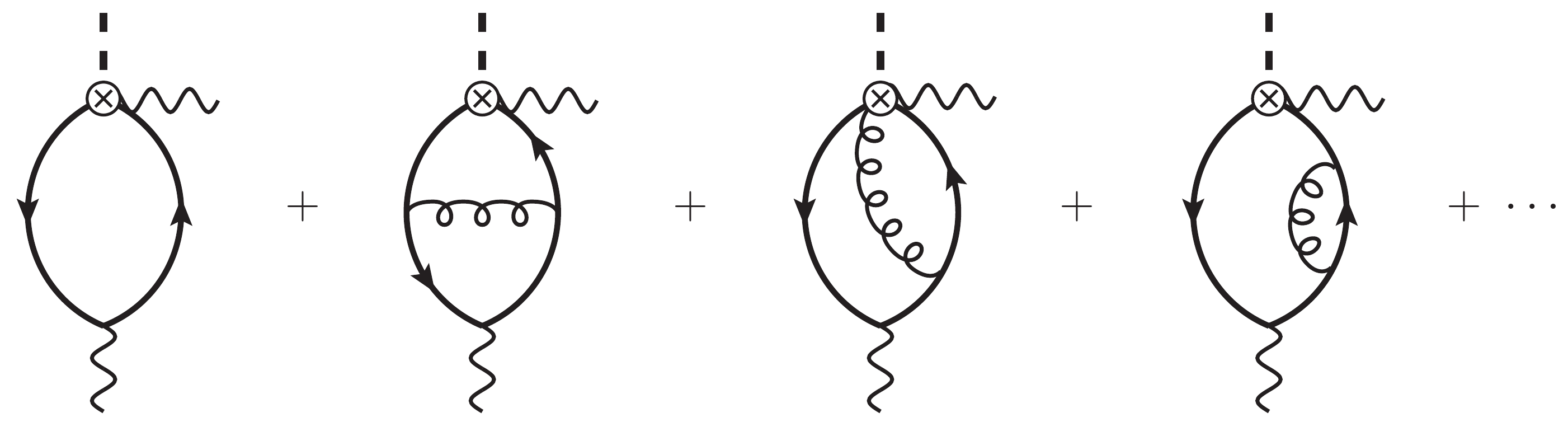}
\vspace{2mm}
\caption{\label{fig:O2graphs} 
Representative Feynman diagrams contributing to the matrix element $\langle\gamma\gamma|\,O_2(z)\,|h\rangle$ up to ${\cal O}(\alpha_s)$. We do not show mirror copies of the last two graphs, in which the gluon is attached to the other side.} 
\end{center}
\end{figure}

For our purposes, it is more important to have the matrix elements in SCET-2 obtained with physical (on-shell) photon states. To all orders in QCD perturbation theory, the matrix element of $O_1$ is given by (omitting the polarization vectors for simplicity) 
\begin{equation}\label{O1bare}
   \langle\gamma\gamma|\,O_1\,|h\rangle
   = m_{b,0}\,g_\perp^{\mu\nu} \,,
\end{equation}
where $m_{b,0}$ is the bare $b$-quark mass. The reason is simply that $O_1$ does not contain any fields with color charges, and hence there are no QCD corrections to the matrix element. The Feynman rule for the collinear photon fields $\A_{n_i,\mu}^\perp$ gives $e_b\,\varepsilon_\mu^{\perp*}(k_i)$. The relevant Feynman diagrams for the matrix element $\langle\gamma\gamma|\,O_2(z)\,|h\rangle$ are depicted in Figure~\ref{fig:O2graphs}. This matrix element is symmetric under the exchange of $z$ and $(1-z)$. We have not been able to calculate it in closed form for arbitrary values of $\epsilon$. Expanding the NLO correction in powers of $\epsilon$, we obtain (with $0\le z\le 1$)
\begin{equation}\label{O2mel}
   \langle\gamma\gamma|\,O_2(z)\,|h\rangle 
   = \frac{N_c\alpha_{b,0}}{2\pi}\,m_{b,0}\,g_\perp^{\mu\nu}
    \left[ e^{\epsilon\gamma_E}\,\Gamma(\epsilon)\,\big(m_{b,0}^2\big)^{-\epsilon} 
    + \frac{C_F\alpha_{s,0}}{4\pi}\,\big( m_{b,0}^2 \big)^{-2\epsilon} \big[ K(z) + K(1-z) \big] \right] ,
\end{equation}
where
\begin{equation}\label{Kfun}
\begin{aligned}
   K(z) &= \frac{1}{\epsilon^2} \left( \ln z + \frac{3}{2} \right)
    + \frac{1}{\epsilon} \left( \frac{\ln^2 z}{2} - \ln z\ln (1-z) - \frac{1}{4} - \frac{\pi^2}{6} \right) \\
   &\quad\mbox{}+ 6\,\text{Li}_3(z) + \left( 1 - 2z - 2\ln z \right) \text{Li}_2(z) 
    + \frac{\ln^3 z}{6} + \big[ z + \ln(1-z) \big] \ln^2 z \\
   &\quad\mbox{}+ \left( 2\,\text{Li}_2(1-z) - \frac{1}{2} \ln(1-z) - \frac{1+3z}{2} - \frac{\pi^2}{6} \right) \ln z    
    + \frac{3}{2} + \frac{\pi^2}{6} - 4\zeta_3 + {\cal O}(\epsilon) \,.
\end{aligned}
\end{equation}
It is obvious from this result that at NLO in $\alpha_s$ the matrix element contains terms that are singular at $z\to 0$ or $z\to 1$. The former terms are contained in $K(z)$, while the latter ones are contained in $K(1-z)$. In order to compute the convolution of $H_2$ with $O_2$, it is important to calculate these leading singularities exactly, without performing an expansion in $\epsilon$. In the form (\ref{rewrite}), we only need to work out the behavior near $z=0$. We obtain
\begin{equation}
\begin{aligned}
   \lim_{z\to 0} \big[ K(z) + K(1-z) \big]
   &= \frac{e^{2\epsilon\gamma_E}}{1-2\epsilon}\,\bigg[ 2(2-3\epsilon+2\epsilon^2)\,\Gamma^2(\epsilon)
    + 2(1-\epsilon)\,\Gamma(\epsilon)\,\Gamma(2\epsilon)\,\Gamma(-\epsilon) \\
   &\hspace{18.5mm}\mbox{}+ z^\epsilon\,(2-4\epsilon-\epsilon^2)\,
    \frac{\Gamma(2\epsilon)\,\Gamma^2(-\epsilon)}{\Gamma(1-2\epsilon)} \bigg] \,.
\end{aligned}
\end{equation}
Note that, after renormalization of the bare couplings according to (\ref{alpharen}), the matrix element in (\ref{O2mel}) depends on the dimensionless ratio $m_{b,0}^2/\mu^2$.

\begin{figure}[t]
\begin{center}
\includegraphics[width=0.9\textwidth]{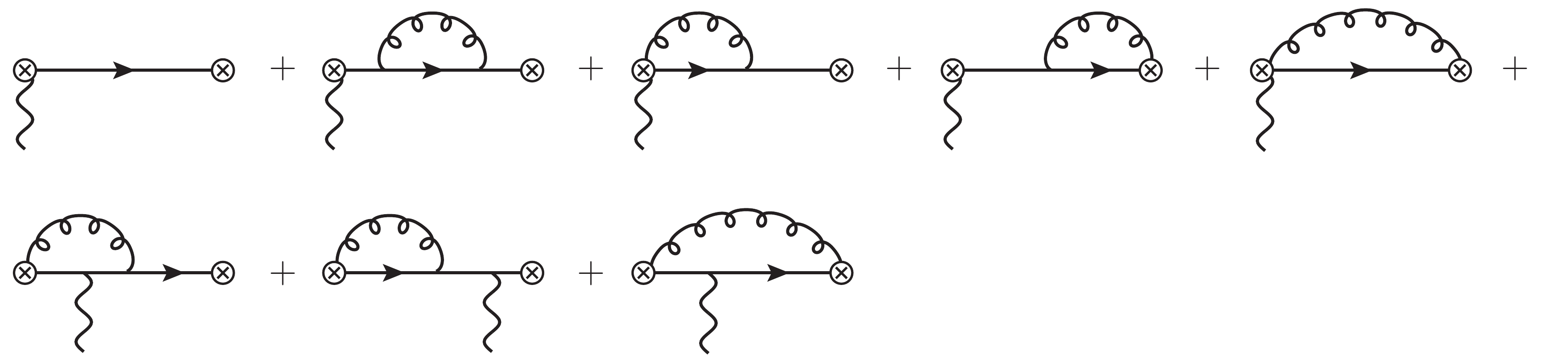}
\vspace{2mm}
\caption{\label{fig:Jgraphs} 
Feynman diagrams contributing to the jet function $J(p^2)$ up to ${\cal O}(\alpha_s)$.} 
\end{center}
\end{figure}

\begin{figure}[t]
\begin{center}
\includegraphics[width=0.65\textwidth]{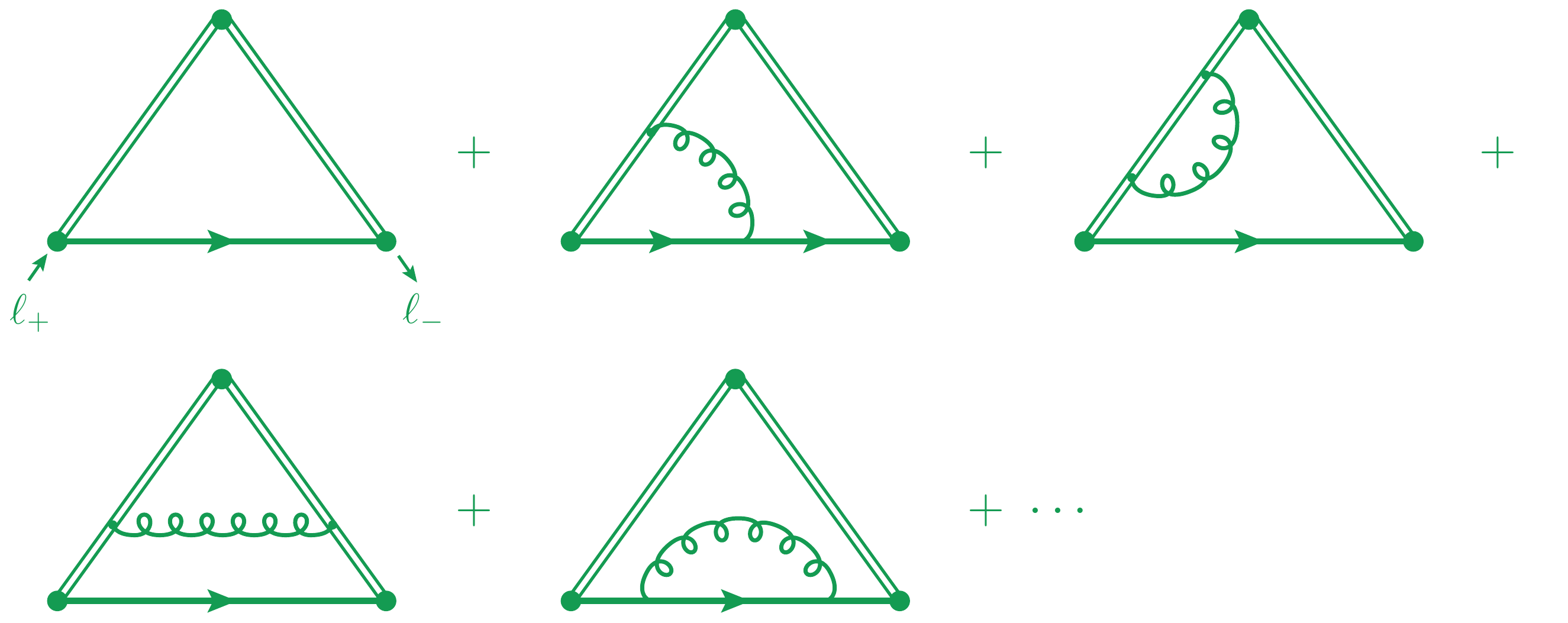}
\vspace{2mm}
\caption{\label{fig:Sgraphs} 
Representative Feynman diagrams contributing to the soft function ${\cal S}(\ell_+\ell_-)$ up to ${\cal O}(\alpha_s)$. We omit the mirror graphs of the second and third diagram, in which the gluon is attached to the other Wilson line. The soft function $S(\ell_+\ell_-)$ in (\ref{SDisc}) is given by the discontinuity of ${\cal S}(\ell_+\ell_-)$.} 
\end{center}
\end{figure}

Concerning the matrix element of the operator $O_3$ in (\ref{Ondef3}), we now present our explicit expressions for the jet and soft functions at NLO in $\alpha_s$. The diagrams contributing to the jet function are shown in Figure~\ref{fig:Jgraphs}. We obtain
\begin{equation}\label{Jfun}
   J(p^2) = 1 + \frac{C_F\alpha_{s,0}}{4\pi} \left( -p^2 -i0 \right)^{-\epsilon}
    e^{\epsilon\gamma_E}\,\frac{\Gamma(1+\epsilon)\,\Gamma^2(-\epsilon)}{\Gamma(2-2\epsilon)}\,
   (2-4\epsilon-\epsilon^2) \,.
\end{equation}
This function has a cut along the positive $p^2$ axis starting at $p^2=0$ and extending to infinity. To the best of our knowledge, this expression (expanded in powers of $\epsilon$) was first obtained in the discussion of the exclusive $B$-meson decay process $B^-\to\gamma l^-\bar\nu_l$ in \cite{Bosch:2003fc}. The calculation of the soft function is more complicated. The relevant Feynman graphs are shown in Figure~\ref{fig:Sgraphs}. Due to the multipole expansion applied to soft fields in interaction terms with hard-collinear fields, the soft momentum component $\ell_+$ enters at the left lower vertex, while $\ell_-$ exists at the right lower vertex, as indicated in the first graph. In a more complicated diagram such as the second graph, the assignment of momenta becomes non-trivial. For example, if the gluon emitted from the Wilson-line segment belongs to $S_{n_1}^{(G)\dagger}(0)$ in (\ref{eq25}), then the plus component of the momentum flowing through the left quark propagator is equal to $\ell_+$, while the minus component of the momentum carried by the right quark propagator is equal to $\ell_-$. It follows that the left quark propagator carries the minus component $(\ell_- -p_-)$, while the right one carries the plus component $(\ell_+ + p_+)$, where $p$ denotes the gluon momentum. As a result, we find that the soft function $S(\ell_+\ell_-)$ defined via the discontinuity of the diagrams shown in Figure~\ref{fig:Sgraphs} has support for all values $\ell_+\ell_- >0$, even though at leading order (first graph) the discontinuity arises only if $\ell_+\ell_->m_{b,0}^2$. Writing
\begin{equation}\label{Ssplit}
   S(\ell_+\ell_-) 
   = - \frac{N_c\alpha_{b,0}}{\pi}\,m_{b,0}\,\Big[ S_a(\ell_+\ell_-)\,\theta(\ell_+\ell_- -m_{b,0}^2)
    + S_b(\ell_+\ell_-)\,\theta(m_{b,0}^2-\ell_+\ell_-) \Big] \,,
\end{equation}
we find (with $w=\ell_+\ell_-$)
\begin{equation}\label{SaSb}
\begin{aligned}
   S_a(w) &= \frac{e^{\epsilon\gamma_E}}{\Gamma(1-\epsilon)}\,\big(w-m_{b,0}^2\big)^{-\epsilon} \\
   &\mbox{}+ \frac{C_F\alpha_{s,0}}{4\pi} \bigg[ 
    \Big[ C_1(\epsilon) + \frac{2}{\epsilon}\,\ln(1-r) \Big] \big(w-m_{b,0}^2\big)^{-2\epsilon} 
    + C_2(\epsilon) \big( m_{b,0}^2 \big)^{1-\epsilon} \big(w-m_{b,0}^2 \big)^{-1-\epsilon} \\
   &\hspace{2.0cm}\mbox{}- 2\,\text{Li}_2(r) + 2 \ln r \ln(1-r) - 3 \ln^2(1-r) + 2 \ln(1-r) + \dots \bigg] \,, \\
   S_b(w) &= \frac{C_F\alpha_{s,0}}{4\pi}\,\big( m_{b,0}^2 \big)^{-2\epsilon}\,\bigg[ 
    - \frac{4}{\epsilon}\,\ln(1-\hat w) + 6 \ln^2(1-\hat w) + \dots \bigg] \,,
\end{aligned}
\end{equation}
where
\begin{equation}
\begin{aligned}
   C_1(\epsilon) &= \frac{2 e^{2\epsilon\gamma_E}}{\Gamma(1-2\epsilon)}
    \left[\frac{(1+\epsilon)\,\Gamma(-\epsilon)^2}{\Gamma(2-2\epsilon)}
    + 2\Gamma(\epsilon)\,\Gamma(-\epsilon) \right] , \\
   C_2(\epsilon) &= - 2 e^{2\epsilon\gamma_E}\,\frac{3-2\epsilon}{1-2\epsilon}\,
    \frac{\Gamma(\epsilon)}{\Gamma(-\epsilon)} \,.
\end{aligned}
\end{equation}
In (\ref{SaSb}) we have defined the dimensionless ratios $r=m_{b,0}^2/w$ and $\hat w=w/m_{b,0}^2$, both of which live on the interval $[0,1]$. We have obtained the exact functional dependence on $r$ and $\hat w$ in terms of generalized hypergeometric functions ${}_2F_1(\dots)$. The resulting expressions are collected in Appendix~\ref{app:soft}. Here it suffices to quote the answer in term of series expansions in $\epsilon$. In both expressions the dots refer to terms of ${\cal O}(\epsilon)$ and higher, which vanish for $r\to 0$ or $\hat w\to 0$, respectively, and which are therefore not needed to reproduce the amplitude at NLO in $\alpha_s$. Note that the factor $\alpha_{b,0}$ in (\ref{Ssplit}) ensures that after coupling renormalization the soft function depends on the dimensionless ratios $m_{b,0}^2/\mu^2$, $r$ and~$\hat w$.

We conclude this section with an important remark. The term proportional to $C_2(\epsilon)$ in (\ref{SaSb}) exhibits a stronger singularity at $w=m_{b,0}^2$ than the remaining terms. However, this term is absorbed entirely when the bare mass parameter $m_{b,0}$ inside the leading-order term in $S_a(w)$ is renormalized and replaced by the $b$-quark pole mass.

\section{Rapidity divergences and analytic regulators}
\label{sec:rapidity}

The unregularized endpoint divergences of the convolution integrals involving the operators $O_2$ and $O_3$ can be tamed by introducing a rapidity regulator. Since the $h\to\gamma\gamma$ process involves time-like kinematics, however, none of the regulators suggested in the literature can be used consistently in our case. Regularizing gluon emissions from Wilson lines \cite{Chiu:2012ir} is not sufficient in our case, since the divergences arise already at zeroth order in $\alpha_s$. Using a non-analytic regulator such as $|2\ell_z|^\eta$, as proposed in \cite{Rothstein:2016bsq}, would destroy the analytic properties of the integral involving the soft function in (\ref{Ondef2}), and hence the steps that led to (\ref{Ondef3}) would no longer hold. Also, with this regulator one would loose certain minus signs in the arguments of logarithms, in such a way that the analytic properties of the decay amplitude are not correctly reproduced. In our case the rapidity logarithms in the sum of all terms in the factorization theorem (\ref{fact1}) must conspire to give $\ln(-M_h^2/m_b^2-i0)$ with a non-zero imaginary part.

\subsection{Introducing rapidity regulators}
\label{subsec:5.1}

\begin{wrapfigure}{r}{8.2cm}
\hspace{2mm}\includegraphics[scale=0.4]{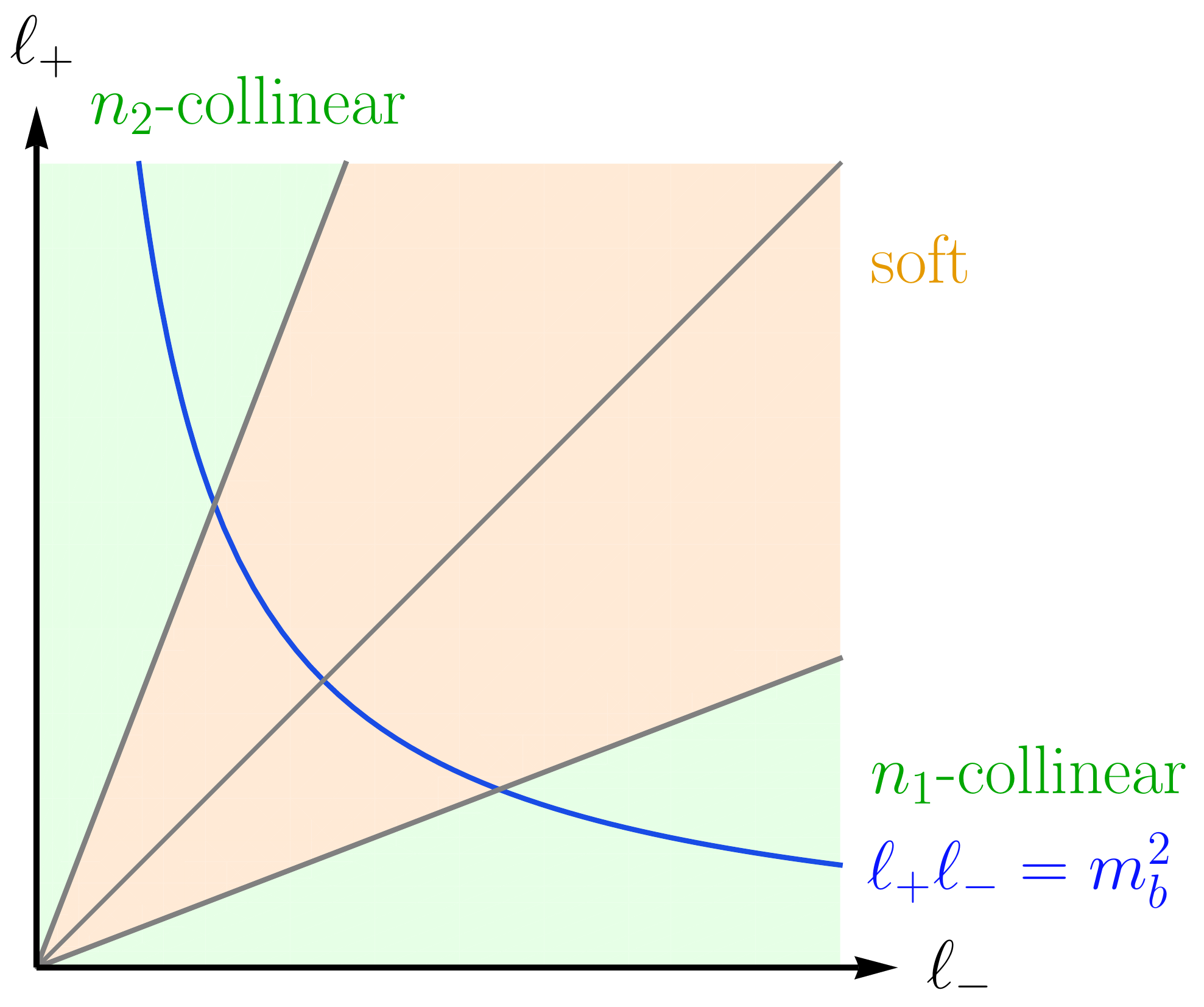}
\caption{\label{fig:5} 
Soft and collinear regions in the plane of the soft momenta $\ell_-$ and $\ell_+$. Contours of constant $\ell_+\ell_-$ are hyperbolas, as indicated by the blue line. The separation of the three regions must be defined using rapidity regulators.}
\end{wrapfigure}

In our work, we will employ an analytic regulator on the convolution variables $z_{1,2}$ and $\ell_\pm$ in the factorization theorems (\ref{Ondef3}) and (\ref{fact1}). Figure~\ref{fig:5} illustrates the physical origin of the endpoint divergences. In the $n_1$-collinear region the variable $z_1=\ell_-/(\bar n_1\cdot k_1)$ is formally of ${\cal O}(1)$, but it is integrated over the interval $z_1\in[0,1]$, and the region near $z_1=0$ gives an unsuppressed contribution. The same applies in the $n_2$-collinear region, where $z_2=\ell_+/(\bar n_2\cdot k_2)$. In the soft region, on the other hand, the variables $\ell_\pm$ are formally of ${\cal O}(m_b)$, but they are integrated up to infinity. We thus see that the collinear and soft regions overlap, and this overlap must be avoided by introducing appropriate regulators. Rather than introducing the regulators in 
the SCET Feynman rules, in our problem it is simplest to impose them on the convolution variables themselves. This is similar in spirit to the approach of \cite{Becher:2011dz}, where the regulator was imposed on the phase-space integrals for real emissions. For the soft contribution, we use the regulator
\begin{equation}
   \left( \frac{-\bar n_1\cdot k_1\,\ell_+ + \bar n_2\cdot k_2\,\ell_- -i0}{\nu^2} \right)^\eta
\end{equation}
under the integral in (\ref{Ondef2}), which gives rise to another branch cut above the real axis in the complex $\ell_-$ plane and thus does not invalidate the arguments that led to (\ref{Ondef3}). Here $\eta$ is an infinitesimal analytic regulator and $\nu$ denotes the associated mass scale. For the collinear contributions, either $\ell_+$ or $\ell_-$ is of ${\cal O}(M_h)$, whereas the other variable is of ${\cal O}(m_b^2/M_h)$. We must therefore multipole expand the regulator term and use 
\begin{equation}
   \left( \frac{-z_i M_h^2-i0}{\nu^2} \right)^\eta
\end{equation}
under the integral in (\ref{rewrite}). Using that $M_h^2=\bar n_1\cdot k_1\,\bar n_2\cdot k_2$, we see that the regulators indeed match up at the boundaries of the soft, $n_1$-collinear and $n_2$-collinear regions.

Introducing the regulators as just described, the bare factorization formula (\ref{fact1}) takes the form 
\begin{equation}\label{fact2}
\begin{aligned}
   {\cal M}_b(h\to\gamma\gamma)
   &= \lim_{\eta\to 0}\,H_1\,\langle\gamma\gamma|\,O_1\,|h\rangle \\
   &\quad\mbox{}+ 2\int_0^1\!\frac{dz_1}{z_1} \left( \frac{-z_1 M_h^2 -i0}{\nu^2} \right)^\eta 
     \bar H_2(z_1)\,\langle\gamma\gamma|\,O_2(z_1)\,|h\rangle + (z_1\to z_2) \\
   &\quad\mbox{}+ \frac{g_\perp^{\mu\nu}}{2}\,H_3 
    \int_0^\infty\!\frac{d\ell_-}{\ell_-} \int_0^\infty\!\frac{d\ell_+}{\ell_+}\,S(\ell_+\ell_-) \\
   &\quad\times \bigg[ \left( \frac{-\bar n_1\cdot k_1\,\ell_+ + \bar n_2\cdot k_2\,\ell_- -i0}{\nu^2} \right)^\eta
     J(\bar n_1\cdot k_1\,\ell_+)\,J(-\bar n_2\cdot k_2\,\ell_-) \\
   &\hspace{1.0cm}\mbox{}+ \left( \frac{\bar n_1\cdot k_1\,\ell_+ - \bar n_2\cdot k_2\,\ell_- -i0}{\nu^2} \right)^\eta 
    J(-\bar n_1\cdot k_1\,\ell_+)\,J(\bar n_2\cdot k_2\,\ell_-) \bigg] \,.
\end{aligned}
\end{equation}
We can now split up the double integral in the last contribution into regions where $\ell_->\ell_+$ and $\ell_+>\ell_-$. In the first region ($\ell_->\ell_+$) rapidity divergences arise for $\ell_-\to\infty$, and they cancel against rapidity divergences of the $n_1$-collinear contribution in the second line, which arise for $z_1=\ell_-/(\bar n_1\cdot k_1)\to 0$. Likewise, in the second region ($\ell_+>\ell_-$) rapidity divergences arise for $\ell_+\to\infty$, and they cancel against rapidity divergences of the $n_2$-collinear contribution in the second line. The two combinations of terms give the same result, so that we can write the final form of the bare factorization theorem as
\begin{equation}\label{fact3}
\begin{aligned}
   {\cal M}_b(h\to\gamma\gamma)
   &= \lim_{\eta\to 0}\,H_1\,\langle\gamma\gamma|\,O_1\,|h\rangle 
    + 4\int_0^1\!\frac{dz}{z} \left( \frac{-z M_h^2 -i0}{\nu^2} \right)^\eta 
     \bar H_2(z)\,\langle\gamma\gamma|\,O_2(z)\,|h\rangle \\
   &\quad\mbox{}+ g_\perp^{\mu\nu}\,H_3 
    \int_0^\infty\!\frac{d\ell_-}{\ell_-} \int_0^{\ell_-}\!\frac{d\ell_+}{\ell_+}\,S(\ell_+\ell_-) \\
   &\quad\times \bigg[ \left( \frac{\bar n_2\cdot k_2\,\ell_- -i0}{\nu^2} \right)^\eta
     J(\bar n_1\cdot k_1\,\ell_+)\,J(-\bar n_2\cdot k_2\,\ell_-) \\
   &\hspace{1.0cm}\mbox{}+ \left( \frac{- \bar n_2\cdot k_2\,\ell_- -i0}{\nu^2} \right)^\eta 
    J(-\bar n_1\cdot k_1\,\ell_+)\,J(\bar n_2\cdot k_2\,\ell_-) \bigg] \,.
\end{aligned}
\end{equation}
In the soft contribution we have dropped the $\ell_+$ components in the regulator terms and kept the larger component $\ell_-$ only. Differences with the original form arise at ${\cal O}(\eta)$ and can be ignored. 

With the regulators in place, the convolution integrals are well defined and can be evaluated in a straightforward way. Endpoint divergences arise from the region $z\to 0$ in the second term and $\ell_-\to\infty$ in the third term. Note that the region $\ell_+\ell_-\to 0$ does not give rise to divergences, because for $\ell_+\ell_-<m_{b,0}^2$ the soft function $S(\ell_+\ell_-)$ is of ${\cal O}(\ell_+\ell_-)$, see (\ref{SaSb}). It is important to perform the expansion in the analytic regulator $\eta$ before the expansion in the dimensional regulator $\epsilon$. Renormalizing the bare $b$-quark Yukawa coupling and quark mass in the $\overline{\rm MS}$ scheme using\footnote{We have factored out $\mu^\epsilon$ from the renormalized Yukawa coupling, so that $y_b(\mu)$ is dimensionless. This factor $\mu^\epsilon$ multiplies the entire decay amplitude and can simply be dropped, because the sum of all contributions is free of $1/\epsilon^n$ poles after the renormalized parameters have been introduced.} 
\begin{equation}\label{myrenorm}
   \frac{y_{b,0}}{\mu^\epsilon\,y_b(\mu)} = \frac{m_{b,0}}{m_b(\mu)} 
   = 1 - \frac{C_F\alpha_s}{4\pi}\,\frac{3}{\epsilon} + {\cal O}(\alpha_s^2) \,,
\end{equation}
we write the answer for the three terms in the factorization formula (\ref{fact3}) in the form
\begin{equation}\label{Tidef}
   {\cal M}_b(h\to\gamma\gamma) = {\cal M}_0 \left( T_1 + T_2 + T_3 \right) ,
\end{equation}
with ${\cal M}_0$ defined in (\ref{M0def}). We find
\begin{equation}\label{HiOires}
\begin{aligned}
   T_1 &= \frac{1}{\epsilon^2} - \frac{L_h}{\epsilon} + \frac{L_h^2}{2} - 2 - \frac{\pi^2}{12} 
    + \frac{C_F\alpha_s}{4\pi}\,k_1(L_h) \,, \\
   T_2 &= \left[ \frac{2}{\eta} + 2 \ln\frac{-M_h^2-i0}{\nu^2} \right] 
    \left[ \frac{1}{\epsilon} - L_m + \frac{C_F\alpha_s}{4\pi}\,k_0(L_h,L_m) \right] 
    + \frac{C_F\alpha_s}{4\pi}\,k_2(L_h,L_m) \,, \\
   T_3 &= - \left[ \frac{2}{\eta} + \ln\frac{-M_h^2-i0}{\nu^2} + \ln\frac{m_b^2}{\nu^2} \right] 
    \left[ \frac{1}{\epsilon} - L_m + \frac{C_F\alpha_s}{4\pi}\,k_0(L_h,L_m) \right] \\
   &\quad\mbox{}- \frac{1}{\epsilon^2} + \frac{L_m}{\epsilon} - \frac{L_m^2}{2} + \frac{\pi^2}{12}
    + \frac{C_F\alpha_s}{4\pi}\,k_3(L_h,L_m) \,, 
\end{aligned}
\end{equation}
where $L_h=\ln(M_h^2/\mu^2)-i\pi$ and $L_m=\ln[m_b(\mu)^2/\mu^2]$. Explicit expressions for the NLO coefficients $k_i$ are listed in Appendix~\ref{app:B}. Adding up the various contributions we recover the leading terms in the decay amplitude given in (\ref{QCDresult}). Note that the $1/\eta$ poles cancel in the sum of all terms, in accordance with the physical picture discussed above. We now discuss the nature of this cancellation in more detail and derive the precise conditions under which it holds.

\subsection{Cancellation of rapidity divergences}

The cancellation of the $1/\eta$ poles in the sum of the contributions $T_2$ and $T_3$ in (\ref{HiOires}) is highly non-trivial, because the first term is a product of functions depending on the hard and collinear scales $-M_h^2$ and $m_b^2$, whereas the latter term also depends on the hard-collinear scale $M_h m_b$. How is it conceivable that such a cancellation could persist in higher orders of perturbation theory? We will show now that this is possible because the hard matching coefficient $\bar H_2(z)$ and the matrix element $\langle\gamma\gamma|\,O_2(z)\,|h\rangle$ in the second term in (\ref{fact3}) can be factorized in the limit $z\to 0$ into products which closely resemble the structure of the third term.\footnote{In the context of exclusive semileptonic $B$-meson decays, such a refactorization of SCET matrix elements in the endpoint region has also been discussed in \cite{Boer:2018}.} 
We introduce the notation $\braces{f(z)}$ to denote all those terms in a function $f(z)$ that are of leading power (modulo logarithms) in the limit $z\to 0$. For the bare functions we will consider, this means that we keep all terms of order $z^{n\epsilon}$ with $n\in\mathbb{Z}$. Using this notation, we find that the required refactorization conditions read
\begin{equation}\label{refact1}
   \braces{\bar H_2(z)} = - H_3\,J(z M_h^2) \,,
\end{equation}
and 
\begin{equation}\label{refact2}
   \braces{\langle\gamma\gamma|\,O_2(z)\,|h\rangle} 
   = - \frac{g_\perp^{\mu\nu}}{2} \int_0^\infty\!\frac{dw}{w}\,J\Big(\!-\frac{w}{z}\Big)\,S(w) 
   = - \frac{g_\perp^{\mu\nu}}{2} \int_0^\infty\!\frac{d\ell_+}{\ell_+}\,
    J(-\bar n_1\cdot k_1\,\ell_+)\,S(\ell_+\ell_-) \,, 
\end{equation}
where we have identified $z=\ell_-/(\bar n_1\cdot k_1)$ and $\ell_+=w/\ell_-$ in the last step. Our explicit expression given in the previous sections confirm that these relations indeed hold at NLO in $\alpha_s$. Concretely, relation (\ref{refact1}) follows directly from (\ref{H2bar}) and (\ref{Jfun}), whereas relation (\ref{refact2}) can be derived from (\ref{Hires}), (\ref{Jfun}) and (\ref{Ssplit}). But indeed these relations must hold to all orders of perturbation theory, and it would be desirable to derive them rigorously from SCET. 

In order to show that the conditions (\ref{refact1}) and (\ref{refact1}) ensure that the rapidity divergences cancel to all orders of perturbation theory, we rewrite the factorization formula (\ref{fact3}) in the short-hand form
\begin{equation}\label{shorthand}
   {\cal M}_b(h\to\gamma\gamma)
   = H_1\,\langle O_1\rangle + 4\bar H_2\otimes\langle O_2\rangle + H_3\,\langle O_3\rangle \,.
\end{equation}
We also adopt the default choice for the reference vectors $n_1$ and $n_2$, so that $\bar n_i\cdot k_i=M_h$ for $i=1,2$. Introducing the new variable $w=\ell_+\ell_-={\cal O}(m_b^2)$ and rearranging the order of integrations, we obtain for the third term
\begin{equation}\label{eq53}
\begin{aligned}
   H_3\,\langle O_3\rangle
   &= g_\perp^{\mu\nu} H_3
    \int_0^\infty\!\frac{dw}{w}\,S(w) \int_{\sqrt{w}}^\infty\frac{d\ell_-}{\ell_-}\, 
    \bigg[ \left( \frac{M_h\ell_-}{\nu^2} \right)^\eta J(M_h w/\ell_-)\,J(-M_h\ell_-) \\
   &\hspace{5.8cm}\mbox{}+ \left( \frac{-M_h\ell_-\!-\!i0}{\nu^2} \right)^\eta J(-M_h w/\ell_-)\,J(M_h\ell_-) \bigg] \,. 
\end{aligned}
\end{equation}
In the second term rapidity divergences arise from the region where $z\to 0$. To evaluate them, we can use the refactorization conditions (\ref{refact1}) and (\ref{refact2}). This leads to
\begin{equation}\label{eq56}
\begin{aligned}
   4\,\braces{\bar H_2}\otimes\braces{\langle O_2\rangle}
   &= 2 g_\perp^{\mu\nu} H_3 \int_0^\infty\!\frac{dw}{w}\,S(w) \int_0^{M_h}\!\frac{d\ell_-}{\ell_-} 
    \left( \frac{-M_h\ell_-\!-\!i0}{\nu^2} \right)^\eta J(-M_h w/\ell_-)\,J(M_h\ell_-) \,.
\end{aligned}
\end{equation}
To proceed, we expand the bare jet function (\ref{Jfun}) in a power series in the bare coupling, 
\begin{equation}
   J(p^2) = \sum_{n=0}^\infty\,c_n(\epsilon) \left( \frac{\alpha_{s,0}}{4\pi} \right)^n
    \left( -p^2-i0 \right)^{-n\epsilon} ,
\end{equation} 
with $c_0(\epsilon)=1$. Using this expansion in (\ref{eq53}), we obtain a double sum over terms proportional to $c_m(\epsilon)\,c_n(\epsilon)$. We find that the terms with $m\ne n$ cancel out in the sum of the two contributions, while those with $m=n$ give
\begin{equation}
\begin{aligned}
   H_3\,\langle O_3\rangle
   &= - g_\perp^{\mu\nu} H_3 \int_0^\infty\!\frac{dw}{w}\,S(w)
    \left[ \frac{2}{\eta} + \ln\frac{-M_h^2-i0}{\nu^2} + \ln\frac{w}{\nu^2} + {\cal O}(\eta) \right] \\
   &\hspace{8mm}\times \sum_{n=0}^\infty\,c_n^2(\epsilon) \left( \frac{\alpha_{s,0}}{4\pi} \right)^{2n}
    \left( - w M_h^2-i0 \right)^{-n\epsilon} \,.
\end{aligned}
\end{equation}
The terms inside the bracket contain the rapidity divergence and rapidity logarithms, whereas the genuine hard-collinear logarithms $\ln(- w M_h^2)$ in the second line appear first at ${\cal O}(\alpha_s^2)$. The corresponding expression derived from (\ref{eq56}) reads
\begin{equation}
\begin{aligned}
   4\,\braces{\bar H_2}\otimes\braces{\langle O_2\rangle}
   &= 2 g_\perp^{\mu\nu} H_3 \int_0^\infty\!\frac{dw}{w}\,S(w)\,\bigg\{
    \left[ \frac{1}{\eta} + \ln\frac{-M_h^2-i0}{\nu^2} + {\cal O}(\eta) \right] \\
   &\hspace{47mm}\times \sum_{n=0}^\infty\,c_n^2(\epsilon) \left( \frac{\alpha_{s,0}}{4\pi} \right)^{2n}
    \left( - w M_h^2-i0 \right)^{-n\epsilon} \\
   &\quad\mbox{}- \sum_{m\ne n}\,\frac{c_m(\epsilon)\,c_n(\epsilon)}{(m-n)\,\epsilon}
    \left( \frac{\alpha_{s,0}}{4\pi} \right)^{m+n}\!
    \left( -M_h^2-i0 \right)^{-m\epsilon} w^{-n\epsilon} \bigg\} \,.
\end{aligned}
\end{equation}
In the sum of the two contributions the $1/\eta$ poles cancel to all orders of perturbation theory, and we obtain 
\begin{equation}\label{eq58}
\begin{aligned}
   &\lim_{\eta\to 0}\,4\,\braces{\bar H_2}\otimes\braces{\langle O_2\rangle} + H_3\,\langle O_3\rangle \\
   &= g_\perp^{\mu\nu} H_3 \int_0^\infty\!\frac{dw}{w}\,S(w)\,\bigg[
    \ln\frac{-M_h^2-i0}{w}\,\sum_{n=0}^\infty\,c_n^2(\epsilon) \left( \frac{\alpha_{s,0}}{4\pi} \right)^{2n}
    \left( - w M_h^2-i0 \right)^{-n\epsilon} \\
   &\hspace{4.25cm}\mbox{}- 2 \sum_{m\ne n}\,\frac{c_m(\epsilon)\,c_n(\epsilon)}{(m-n)\,\epsilon}
    \left( \frac{\alpha_{s,0}}{4\pi} \right)^{m+n}
    \left( -M_h^2-i0 \right)^{-m\epsilon} w^{-n\epsilon} \bigg] \,.
\end{aligned}
\end{equation}

At first sight, the fact that the rapidity divergences cancel between the second and third term in the factorization theorem (\ref{fact2}) appears like a miracle, because these terms depend on different scales. The second term is a convolution of a hard function (depending on the scale $-M_h^2$) and a collinear function (depending on the scale $m_b^2$). The third term, instead, is a product of a hard function (depending on $-M_h^2$), two hard-collinear functions (depending on the scale $M_h m_b$) and a soft function (depending on the scale $m_b^2$). What happens is that in (\ref{eq53}) the two jet functions combine in such a way that their convolutions depend only on the products of scales $(\pm M_h w/\ell_-)(\mp M_h\ell_-)=-w M_h^2$. This is parametrically equal to the product of the hard and collinear scales. This non-trivial fact makes the all-order cancellation possible.

At this point it is instructive to compare our results (\ref{fact3}) and (\ref{eq58}) with the corresponding expressions obtained in the simpler case of Sudakov resummation at leading power, as discussed in detail in \cite{Chiu:2007yn,Chiu:2007dg}. An example would be the decay $Z'\to q\bar q$, where $Z'$ is a color-neutral heavy vector boson. In this case there is no analogue of the operator $O_1$, because the particles produced at the decay vertex of the $Z'$ boson are the same as the particles in the external state. More importantly, however, the are no non-trivial jet functions, because soft gluon exchanges between the (hard-)collinear quarks are described by soft Wilson lines, to all orders of perturbation theory. This leads to drastic simplifications in (\ref{eq58}), as we can see by setting all expansion coefficients $c_n(\epsilon)$ of the jet function to zero, except for $c_0=1$. This leads to
\begin{equation}\label{eq61}
   4\,\braces{\bar H_2}\otimes\braces{\langle O_2\rangle} + H_3\,\langle O_3\rangle 
   \to g_\perp^{\mu\nu} H_3\,\bigg[ \ln\frac{-M_h^2-i0}{m_{b,0}^2} \int_0^\infty\!\frac{dw}{w}\,S(w)
    - \int_0^\infty\!\frac{dw}{w}\,S(w)\,\ln\frac{w}{m_{b,0}^2} \bigg] \,.
\end{equation}
The two integrals over the soft function define two soft matrix elements living at the scale $m_{b,0}^2$. The structure that multiplies the hard matching coefficient $H_3$ then contains two terms: one associated with a {\em single\/} large rapidity logarithm and one without such an enhancement. In other words, the relevant soft matrix element is a linear function of the large rapidity logarithm, in agreement with the findings of \cite{Chiu:2007yn,Chiu:2007dg}.

\section{Subtraction of endpoint divergences}
\label{sec:subtr}

We would now like to find a way to reproduce the answer (\ref{eq58}) from an expression that is structurally similar to the original expression (\ref{fact2}) but free of endpoint divergences. This is accomplished by the rearrangement 
\begin{equation}\label{fact4}
\begin{aligned}
   {\cal M}_b(h\to\gamma\gamma)
   &= \left( H_1 + \Delta H_1 \right) \langle\gamma\gamma|\,O_1\,|h\rangle \\
   &\quad\mbox{}+ 4 \lim_{\delta\to 0}\,\int_\delta^1\!\frac{dz}{z}\,
    \Big[ \bar H_2(z)\,\langle\gamma\gamma|\,O_2(z)\,|h\rangle
    - \braces{\bar H_2(z)}\,\braces{\langle\gamma\gamma|\,O_2(z)\,|h\rangle} \Big] \\
   &\quad\mbox{}+ g_\perp^{\mu\nu} H_3\,\lim_{\sigma\to-1}\, 
    \int_0^{M_h}\!\frac{d\ell_-}{\ell_-}\,\int_0^{\sigma M_h}\!\frac{d\ell_+}{\ell_+}\,
    J(M_h\ell_-)\,J(-M_h\ell_+)\,S(\ell_+\ell_-)\,\Big|_{\rm leading\;power}
\end{aligned}
\end{equation}
of the factorization formula. Note that, by construction, the second term is free of endpoint divergences. The regulator $\delta$ is introduced only in an intermediate step, such that both integrals are well defined. Also, as mentioned earlier, the region where $\ell_+\ell_-\to 0$ in the last term does not give rise to endpoint divergences. The limit $\sigma\to-1$ in this term is meant in the sense of analytic continuation. The integral over $\ell_+$ must be evaluated assuming that the upper limit $\sigma M_h>0$, and the result must then be analytically continued to $\sigma M_h=-M_h-i0$. Note that the term in the third line of (\ref{fact4}) is structurally similar to the resummation approach of \cite{Kotsky:1997rq,Akhoury:2001mz,Penin:2014msa,Liu:2017vkm,Liu:2018czl}, where the cutoffs on the integrals over $\ell_\pm$ were implemented by hand. However, we have provided precise definitions of the hard, jet and soft functions in this term, which are valid to all orders of perturbation theory. Also, we have derived the remaining terms in the factorization formula, which have not been discussed previously in the literature.

The contribution $\Delta H_1$ is a ``hard subtraction'' defined by
\begin{equation}\label{DH1}
\begin{aligned}
   \Delta H_1 
   &= \frac{H_3}{m_{b,0}}\,\lim_{\sigma\to-1}\, \int_{\sigma M_h^2}^\infty\!\frac{dw}{w}\,S_\infty(w)\,
    \bigg\{ \ln\frac{-M_h^2-i0}{w}\,
    \sum_{n=0}^\infty\,c_n^2(\epsilon) \left( \frac{\alpha_{s,0}}{4\pi} \right)^{2n} 
    \left( - w M_h^2-i0 \right)^{-n\epsilon} \\[-1mm]
   &\hspace{5.6cm}\mbox{}- 2 \sum_{m\ne n}\,\frac{c_m(\epsilon)\,c_n(\epsilon)}{(m-n)\,\epsilon}
    \left( \frac{\alpha_{s,0}}{4\pi} \right)^{m+n}
    \left( -M_h^2-i0 \right)^{-m\epsilon} w^{-n\epsilon} \bigg\} \,.
\end{aligned}
\end{equation}
The function $S_\infty(w)$ is obtained by setting $m_{b,0}\to 0$ in the expression for the soft function in (\ref{SaSb}), which leads to
\begin{equation}
   S_\infty(w) 
   = - \frac{N_c\alpha_{b,0}}{\pi}\,m_{b,0}\,\bigg[ \frac{e^{\epsilon\gamma_E}}{\Gamma(1-\epsilon)}\,w^{-\epsilon} 
    + \frac{C_F\alpha_{s,0}}{4\pi}\,C_1(\epsilon)\,w^{-2\epsilon} \bigg] \,.
\end{equation}
As a result, $\Delta H_1$ depends on the hard scale $(-M_h^2)$ only. After a straightforward calculation, we obtain at NLO in $\alpha_s$
\begin{equation}
\begin{aligned}
   \Delta H_1 
   &= - \frac{y_{b,0}}{\sqrt2}\,\frac{N_c\alpha_{b,0}}{\pi} \left( -M_h^2 -i0 \right)^{-\epsilon}
    \frac{e^{\epsilon\gamma_E}}{\epsilon^2\,\Gamma(1-\epsilon)}\,
    \bigg\{ 1 + \frac{C_F\alpha_{s,0}}{4\pi} \left( -M_h^2-i0 \right)^{-\epsilon} \\
   &\qquad\times 
    \frac{e^{\epsilon\gamma_E}\,\Gamma(-\epsilon)\,\Gamma(1-\epsilon)}{\Gamma(2-2\epsilon)} \left[ (1-2\epsilon+3\epsilon^2)\,\Gamma(\epsilon) 
    + \frac{1+\epsilon}{2}\,\frac{\Gamma(-\epsilon)}{\Gamma(1-2\epsilon)} \right] \bigg\} \,.
\end{aligned}
\end{equation}

Note that the third contribution in (\ref{fact4}) contains some power-suppressed terms of order $m_b^2/M_h^2$ arising from the upper limits of the integrations over $\ell_\pm$. These must be dropped for consistency. In the definition of $\Delta H_1$, we have omitted such power corrections by using the limiting function $S_\infty(w)$ instead of $S(w)$ in (\ref{DH1}). The projection onto the leading-power terms can be accomplished in a similar way be writing
\begin{equation}
\begin{aligned}
    &\hspace{5mm} \int_0^{M_h}\!\frac{d\ell_-}{\ell_-}\,\int_0^{\sigma M_h}\!\frac{d\ell_+}{\ell_+}\,
     J(M_h\ell_-)\,J(-M_h\ell_+)\,S(\ell_+\ell_-)\,\Big|_{\rm leading\;power} \\
    &= \int_0^{\sigma M_h^2} \frac{dw}{w}\,S(w) \int_{\frac{w}{\sigma M_h}}^{M_h}\!\frac{d\ell_-}{\ell_-}\,
     J(M_h\ell_-)\,J(-M_h w/\ell_-)\,\Big|_{\rm leading\;power} \\
    &= \left[ \int_0^{m_{b,0}^2} \frac{dw}{w}\,S(w) 
     + \int_{m_{b,0}^2}^{\sigma M_h^2} \frac{dw}{w}\,S_\infty(w)
     + \int_{m_{b,0}^2}^\infty \frac{dw}{w}\,\big[ S(w) - S_\infty(w) \big] \right] \\
    &\quad\times \int_{\frac{w}{\sigma M_h}}^{M_h}\!\frac{d\ell_-}{\ell_-}\,J(M_h\ell_-)\,J(-M_h w/\ell_-) \,.
\end{aligned}
\end{equation}
The last expression is free of power-suppressed terms.

We now obtain for the three terms in the rearranged factorization theorem (\ref{fact4}), written as in (\ref{Tidef}),
\begin{equation}
\begin{aligned}
   T_1 &= - 2 + \frac{C_F\alpha_s}{4\pi}\,\bigg[ - \frac{\pi^2}{3\epsilon^2} 
    + \frac{1}{\epsilon} \left( \frac{2\pi^2}{3}\,L_h - 10\zeta_3 \right) 
    - \frac{2\pi^2}{3}\,L_h^2 + \left( 12 + 20\zeta_3 \right) L_h - 36 - \frac{7\pi^4}{30} \bigg] \,, \\
   T_2 &= \frac{C_F\alpha_s}{4\pi}\,\bigg[ \frac{\pi^2}{3\epsilon^2} 
    + \frac{1}{\epsilon} \left( - \frac{2\pi^2}{3}\,L_h + 2\zeta_3 \right) 
    + \frac{\pi^2}{3} \left( L_h^2 + 2 L_h L_m - L_m^2 \right) - 4\zeta_3\,L_h + 8\zeta_3 
    + \frac{13\pi^4}{90} \bigg] \,, \\
   T_3 &= \frac{L^2}{2} + \frac{C_F\alpha_s}{4\pi}\,\bigg[ \frac{8\zeta_3}{\epsilon} - \frac{L^4}{12} - L^3 
    + \left( 4 - \frac{\pi^2}{3} \right) L^2 - \left( 8 - \frac{2\pi^2}{3} \right) L \\
   &\hspace{2.85cm}\mbox{}+ L_m \left( - 3 L^2 + 6 L - 16\zeta_3 \right) - 4\zeta_3 - \frac{\pi^4}{9} \bigg] \,, 
\end{aligned}
\end{equation}
where in the last term $L=L_h-L_m$. Adding up the three contributions we correctly reproduce the QCD amplitude in (\ref{QCDresult}). Note that the sum of the first two terms $(T_1+T_2)$ as well as the third term $T_3$ by itself are ``almost'' RG invariant. We believe that the remaining $\zeta_3/\epsilon$ terms are removed when the bare component functions in the factorization theorem are expressed in terms of renormalized functions.

Some comments are in order concerning the structure of the various terms in (\ref{fact4}). The subtraction of the leading singular terms in the second line, which removes the endpoint divergences of the $z$ integral, generalizes a subtraction prescription proposed in \cite{Alte:2018nbn,Beneke:2019kgv}, which works only for cases where the hard matching coefficient corresponding to $\bar H_2(z)$ approaches a constant plus power-suppressed terms as $z\to 0$. In the present case the situation is more complicated, because $\bar H_2(z)$ contains powers of $\ln z$ and hence is more singular. The form of the third term in (\ref{fact4}) appears to violate strict factorization, because the soft momentum components $\ell_+$ and $\ell_-$ scale like $m_b$ in the soft region, but the integrals over these components extend into the hard region. In other word, this third term contains large (rapidity) logarithms beyond those arising from the RG evolution of the various component functions. This is a result of the collinear anomaly \cite{Becher:2010tm}, which provides an extra source of large logarithms (so-called rapidity logarithms \cite{Chiu:2012ir}) not related to conventional scale evolution. In simpler cases such as the $q_T$ resummation in Drell-Yan \cite{Becher:2011xn} or Higgs production \cite{Becher:2012yn}, the collinear anomaly introduces a dependence of the product of two collinear beam functions on the hard scale, thereby violating strict factorization. Using simple differential equations, one can show that to all orders of perturbation theory this additional source of large logarithms takes the form of a linear function of $\ln(Q^2/q_T^2)$ in the logarithm of the cross section. In the present case, the anomalous dependence on the hard scale enters via the upper limits on the integrals in the third term in (\ref{fact4}) and is thus of a more complicated nature.

\section{Resummation of the leading double logarithms}
\label{sec:6}

In (\ref{fact4}) we have obtained a factorization theorem for the $b$-quark induced $h\to\gamma\gamma$ decay amplitude which is free of endpoint divergences. The collinear anomaly provides a source of large rapidity logarithms, which are contained in the third term. They arise from the fact that the soft momentum components $\ell_\pm$ are integrated up to values of ${\cal O}(M_h)$. Nevertheless, these logarithms would be under control if one knew how to obtain reliable results for the jet and soft functions $J(p^2)$ and $S(w)$ for arbitrary values of their arguments.

In its present form, the factorization theorem is written in terms of bare Wilson coefficients and operator matrix elements. The next step would be to replace these objects in terms of renormalized component functions. Note that except for the hard function $H_3$ all quantities must be renormalized ``non-locally'', by integrating them with $Z$-factors depending on two variables $z$ and $z'$ or $\ell_\pm$ and $\ell_\pm'$. There is also a non-trivial operator mixing, such that the operators $O_2$ and $O_3$ require counterterms involving $O_1$. While we have been able to calculate the relevant renormalization factors for the first two terms in the factorization theorem (including their mixing), the renormalization factors for the jet and soft functions are currently still unknown. We thus leave a detailed discussion of renormalization for future work.

The situation simplifies drastically if we are only interested in the leading double-logarith\-mic corrections to the decay amplitude. They arise from the $\alpha_s/\epsilon^2$ pole terms in the expressions of the bare functions contributing to the third term in the factorization formula (\ref{fact4}), because only this term contains an extra enhancement by two rapidity logarithms. These leading poles can readily be identified using our results for the bare functions $H_3$, $J$ and $S$ given in (\ref{Hires}), (\ref{Jfun}), (\ref{Ssplit}) and (\ref{SaSb}). Note that the leading poles $(\alpha_s/\epsilon^2)^n$ in higher orders are all governed by the same coefficients. After renormalization of the bare quark mass and Yukawa coupling, we obtain 
\begin{equation}\label{doublepoles}
\begin{aligned}
   H_3 &= - \frac{y_b}{\sqrt2} \left[ 1 + \frac{C_F\alpha_s}{4\pi}
    \left( - \frac{2}{\epsilon^2} + \frac{2}{\epsilon}\,\ln\frac{-M_h^2-i0}{\mu^2} 
    - \ln^2\frac{-M_h^2-i0}{\mu^2} + \dots \right) \right] , \\
   J(p^2) &= 1 + \frac{C_F\alpha_s}{4\pi}
    \left( \frac{2}{\epsilon^2} - \frac{2}{\epsilon}\,\ln\frac{-p^2-i0}{\mu^2} 
    + \ln^2\frac{-p^2-i0}{\mu^2} + \dots \right) , \\
   S(w) &= - \frac{N_c\alpha_b}{\pi}\,\frac{m_b\,e^{\epsilon\gamma_E}}{\Gamma(1-\epsilon)}
    \big( w-m_b^2 \big)^{-\epsilon}\,\theta(w-m_b^2)\! \left[ 1 + \frac{C_F\alpha_s}{4\pi}
    \left(\! - \frac{2}{\epsilon^2} + \frac{2}{\epsilon}\,\ln\frac{w}{\mu^2} 
    - \ln^2\frac{w}{\mu^2} + \dots \right) \! \right] \!,
\end{aligned}
\end{equation}
where the dots refer to terms which do not contribute to the leading double logarithms. The choice of the argument in the logarithms of the soft function is not unique, since this function depends on both $m_b^2$ and $w$. Our choice of using the ratio $w/\mu^2$ is motivated by the fact that in this way the $1/\epsilon^n$ pole terms cancel in the product $H_3\,J(M_h\ell_-)\,J(-M_h\ell_+)\,S(\ell_+\ell_-)$ of the four functions in (\ref{fact4}). What remains is a large double logarithm, 
\begin{equation}\label{eq68}
\begin{aligned}
   & H_3\,J(M_h\ell_-)\,J(-M_h\ell_+)\,S(\ell_+\ell_-) \\
   &= \frac{N_c\alpha_b}{\pi}\,\frac{y_b}{\sqrt2}\,m_b\,\theta(\ell_+\ell_- - m_b^2)
    \left[ 1 - \frac{C_F\alpha_s}{2\pi}\,\ln\frac{M_h}{\ell_-}\,\ln\frac{-M_h-i0}{\ell_+} + \dots \right] .
\end{aligned}
\end{equation}

The leading double poles and the associated single poles multiplying the so-called cusp logarithms of the corresponding hard, hard-collinear and soft scales in (\ref{doublepoles}) can be subtracted by means of local counterterms. The corresponding anomalous dimensions governing the scale evolution of the renormalized functions are of the form
\begin{equation}\label{RGEs}
\begin{aligned}
   \frac{d}{d\ln\mu}\,H_3(\mu) 
   &= \left[ \Gamma_{\rm cusp}(\alpha_s)\,\ln\frac{-M_h^2}{\mu^2} + \dots \right] H_3(\mu) \,, \\
   \frac{d}{d\ln\mu}\,J(M_h\ell_-,\mu)
   &= \left[ - \Gamma_{\rm cusp}(\alpha_s)\,\ln\frac{-M_h\ell_-}{\mu^2} + \dots \right] J(M_h\ell_-,\mu) \,, \\
   \frac{d}{d\ln\mu}\,J(-M_h\ell_+,\mu)
   &= \left[ - \Gamma_{\rm cusp}(\alpha_s)\,\ln\frac{M_h\ell_+}{\mu^2} + \dots \right] J(-M_h\ell_+,\mu) \,, \\
   \frac{d}{d\ln\mu}\,S(\ell_+\ell_-,\mu)
   &= \left[ \Gamma_{\rm cusp}(\alpha_s)\,\ln\frac{\ell_+\ell_-}{\mu^2} + \dots \right] S(\ell_+\ell_-,\mu) \,.
\end{aligned}
\end{equation}
where the dots refer to terms without logarithmic enhancement, and 
\begin{equation}
   \Gamma_{\rm cusp}(\alpha_s)
   = \frac{C_F\alpha_s}{\pi} \left\{ 1 + \left[ \left( \frac{67}{36} - \frac{\pi^2}{12} \right) C_A
    - \frac59\,T_F\,n_f \right] \frac{\alpha_s}{\pi} + {\cal O}(\alpha_s^2) \right\}
\end{equation}
is the cusp anomalous dimension \cite{Korchemsky:1992xv}. Here $n_f=5$ is the number of light quark flavors. To resum the leading double logarithms to all orders of perturbation theory, we solve the above evolution equations by ignoring all but the leading logarithmic terms. At the natural scales for each function, we use the lowest-order matching conditions, i.e.\ $H_3(\mu)=-y_b/\sqrt{2}$ at $\mu^2=-M_h^2$, $J(p^2,\mu)=1$ at $\mu^2=-p^2$, and $S(w,\mu)=-\frac{N_c\alpha_b}{\pi}\,m_b\,\theta(w-m_b^2)$ at $\mu^2=w$. In the approximation where the running of the strong coupling is ignored, and where one uses the one-loop expression for the cusp anomalous dimension, this leads to 
\begin{equation}\label{approxsimple}
\begin{aligned}
   & H_3(\mu)\,J(M_h\ell_-,\mu)\,J(-M_h\ell_+,\mu)\,S(\ell_+\ell_-,\mu) \\
   &= \frac{N_c\alpha_b}{\pi}\,\frac{y_b}{\sqrt2}\,m_b\,\theta(\ell_+\ell_- - m_b^2)\, 
    \exp\!\bigg[ - \frac{C_F\alpha_s}{2\pi}\,\ln\frac{M_h}{\ell_-}\,\ln\frac{-M_h-i0}{\ell_+} + \dots \bigg] \,,
\end{aligned}
\end{equation}
corresponding to a simple exponentiation of the correction given in (\ref{eq68}). Note that the resummation shuts itself off when the soft momentum component $\ell_-$ approaches $M_h$ or $\ell_+$ approaches $-M_h$. Thus, the exponential is well behaved even outside the region of soft momenta $\ell_\pm$. 

Using the approximation (\ref{approxsimple}) in the third term in the factorization formula (\ref{fact4}), we can compute the infinite tower of leading double logarithms by performing the integral 
\begin{equation}\label{Imaster}
\begin{aligned}
   {\cal M}_b(h\to\gamma\gamma) \big|_{\rm LDL}
   &= \!\lim_{\sigma\to -1} {\cal M}_0 \int_0^{\sigma M_h}\!\frac{d\ell_+}{\ell_+} 
    \int_0^{M_h}\!\frac{d\ell_-}{\ell_-}\,\theta(\ell_+\ell_- - m_b^2)\,    
    \exp\!\bigg[ - \frac{C_F\alpha_s}{2\pi} \ln\frac{\sigma M_h}{\ell_+} \ln\frac{M_h}{\ell_-} \bigg] \\
   &= {\cal M}_0 \int_0^L\!dL_1 \int_0^L\!dL_2\,\theta(L-L_1-L_2)\,
    \exp\!\bigg[ - \frac{C_F\alpha_s}{2\pi}\,L_1 L_2 \bigg] \\
   &= {\cal M}_0\,\frac{L^2}{2}\,{}_2F_2\bigg(1,1;\frac32,2;-\frac{C_F\alpha_s}{8\pi}\,L^2\bigg) \,,
\end{aligned}
\end{equation}
where $L=\ln(M_h^2/m_b^2)-i\pi$. This reproduces the result (\ref{Penin}), apart from the fact that in our case the argument of the hypergeometric function involves the correct logarithm $L$, whereas in (\ref{Penin}) the imaginary part ``$-i\pi$'' is missing. From a numerical point of view, including the $i\pi$ terms is more important than resumming the large logarithms beyond NLO in $\alpha_s$. In Table~\ref{tab:numerics}, we present numerical results for the leading double-logarithmic corrections to the decay amplitude in units of ${\cal M}_0$. We use $m_b(m_b)=4.18$\,GeV in the argument of the logarithm and vary the renormalization scale of the 3-loop running coupling $\alpha_s(\mu)$ between $\mu=m_b$ and $\mu=M_h$. The scale dependence gives an indication about the expected impact of higher-order logarithmic corrections.

\begin{table}
\begin{center}
\begin{tabular}{c|ccc|ccc} 
 & \multicolumn{3}{c|}{${\cal M}_b(h\to\gamma\gamma)/{\cal M}_0$}
 & \multicolumn{3}{c}{$|{\cal M}_b(h\to\gamma\gamma)/{\cal M}_0|$} \\
$\mu$ & NLO & Eq.\,(\ref{Penin}) & Eq.\,(\ref{Imaster}) & NLO & Eq.\,(\ref{Penin}) & Eq.\,(\ref{Imaster}) \\
\hline
 $m_b$ & $19.2-15.1 i$ & 19.4 & $18.4-15.8\,i$ & 24.4 & 19.4 & 24.2 \\
 $\sqrt{M_h m_b}$ & $18.8-17.2 i$ & 20.5 & $18.5-17.5 i$ & 25.5 & 20.5 & 25.4 \\
 $M_h$ & $18.7-18.3 i$ & 21.1 & $18.5-18.4 i$ & 26.1 & 21.1 & 26.1 \\
\end{tabular}
\caption{\label{tab:numerics} 
Numerical results for the leading double-logarithmic contributions to the decay amplitude at NLO and after resummation using the expressions given in (\ref{Penin}) \cite{Kotsky:1997rq,Akhoury:2001mz,Liu:2017vkm} and our result (\ref{Imaster}). The three right columns show the corresponding results for the absolute value of the decay amplitude. The renormalization scale of the running coupling $\alpha_s(\mu)$ is varied between $m_b=4.18$\,GeV and $M_h=125.1$\,GeV.}
\end{center}
\end{table}

It is a straightforward task to include the running of the coupling in the solution of the evolution equations (\ref{RGEs}). Defining the Sudakov exponent $S(\mu_0^2,\mu^2)$ (not to be confused with the soft function) as \cite{Neubert:2004dd}
\begin{equation}
   S(\mu_0^2,\mu^2) 
   = - \frac14\,\int_{\mu^2}^{\mu_0^2}\frac{d\nu^2}{\nu^2}\,\Gamma_{\rm cusp}\big(\alpha_s(\nu^2)\big)\,
    \ln\frac{\mu_0^2}{\nu^2} \,,
\end{equation}
we obtain the solution
\begin{equation}\label{eq74}
\begin{aligned}
   & H_3\,J(M_h\ell_-)\,J(-M_h\ell_+)\,S(\ell_+\ell_-)
    = \frac{N_c\alpha_b}{\pi}\,\frac{y_b}{\sqrt2}\,m_b\,\theta(\ell_+\ell_- - m_b^2) \\
   &\times \exp\!\big[ 2S(-M_h^2,\mu^2) - 2S(-M_h\ell_-,\mu^2) - 2S(M_h\ell_+,\mu^2) + 2S(\ell_+\ell_-,\mu^2) \big] \,.
\end{aligned}
\end{equation}
Despite appearance, this expression is independent of the reference scale $\mu^2$. Once again, the resummation shuts itself off when the soft momenta $\ell_\pm$ approach their upper values $\mp M_h$. Note that some terms in (\ref{eq74}) involve the running coupling $\alpha_s(\mu^2)$ evaluated at time-like argument $\mu^2=-|\mu^2|-i0$. This quantity is well defined and can be computed using standard expressions for $\alpha_s(\mu^2)$. Indeed, using Sudakov exponents with time-like arguments is a well-known method to consistently resum large logarithms along with $i\pi$ terms to all orders of perturbation theory \cite{Ahrens:2008qu,Ahrens:2008nc}. The resummation of the leading logarithms including the running of $\alpha_s$ can be obtained by using expression (\ref{eq74}) instead of (\ref{eq68}) under the integral in (\ref{Imaster}). We defer a detailed numerical discussion of resummation effects to future work, where we will also resum large logarithms beyond the leading double-logarithmic approximation.

\section{Conclusions and outlook}
\label{sec:7}

In this work we have started a detailed discussion of factorization at subleading power in SCET. At this order, factorization theorems for high-energy cross sections and decay amplitudes contain endpoint-divergent convolution integrals. Their presence indicates a violation of simple scale separation. These endpoint divergences cannot be removed using standard techniques of dimensional regularization and $\overline{\rm MS}$ subtractions. They thus hint at an unexpected failure of conventional techniques used in renormalization theory. 

With the example of the $b$-quark induced radiative decay $h\to\gamma\gamma$ of the Higgs boson we have worked out a rather complicated example, in which endpoint-divergent convolution integrals give rise to additional singularities that require both the dimensional regulator $\epsilon$ and a rapidity regulator $\eta$. We have derived a factorization theorem for the decay amplitude in terms of bare Wilson coefficients and operator matrix elements, and we have introduced a new rapidity regulator which preserves the analytic properties of the decay amplitude, taking into account the time-like kinematics of the $h\to\gamma\gamma$ process. We have derived the conditions (\ref{refact1}) and (\ref{refact2}) under which the endpoint divergences caused by rapidity divergences cancel to all orders of perturbation theory. Moreover, we have shown that endpoint divergences that are regularized dimensionally can be avoided by rearranging the terms in the factorization theorem. While we have not yet derived the full set of renormalization factors and anomalous dimensions for the various objects entering the factorization theorem, we have succeeded in resumming the leading double-logarithmic corrections of order $\alpha_s^n\ln^{2n+2}(-M_h^2/m_b^2)$ to all orders of perturbation theory.

Our findings have important consequences for the general theory of electroweak Sudakov resummation in SCET. In the standard approach to this problem, as formulated in \cite{Chiu:2007yn,Chiu:2007dg}, one matches the full theory (the SM or an extension thereof) to SCET at a high scale $Q\gg v$, where $v\approx 246$\,GeV (the equivalent of $m_b$ in our approach) denotes the scale of electroweak symmetry breaking, and $Q$ (the equivalent of $M_h$ in our case) can for instance be the large center-of-mass energy of a collider process or the mass of a new heavy particle in a decay process. Evolving the effective theory from the high scale $Q$ to the low scale $v$, one can resum large (double) logarithms of the scale ratio $Q/v\gg 1$. In this evolution the particles of the SM are treated as massless. At a low scale of ${\cal O}(v)$, one then switches to an effective theory with massive SM particles (and broken electroweak symmetry) by performing a second matching step. It was shown in \cite{Chiu:2007yn,Chiu:2007dg} that the SCET matrix elements in this low-energy effective theory contain one power of the large rapidity logarithm $\ln(Q/v)$ due to the collinear anomaly, but that this does not invalidate the resummation procedure just described. Thus, the problem can be treated as a relatively simple two-scale process. We indeed find that this procedure is correct for processes described at leading power in the effective theory; see relation (\ref{eq61}) and the discussion surrounding it. However, it needs to be {\em generalized\/} for processes such as $h\to\gamma\gamma$, which arise at subleading order in the expansion in scale ratios. These processes are characterized by three hierarchical mass scales: the hard scale $Q$, the low scale $v$, and the intermediate hard-collinear scale $\sqrt{Q\hspace{0.5mm}v}$. As can be seen from our result (\ref{eq58}), the hard-collinear scale enters in an essential way, and the consistent resummation of all large logarithms requires that one takes this intermediate scale properly into account.

We close this paper with a speculation. Our final form of the factorization formula in (\ref{fact4}) is explicit and well defined, however it is structurally rather different from the original form of the bare factorization theorem as given in (\ref{Ondef3}) and (\ref{fact1}). On the other hand, we have noted earlier in Section~\ref{subsec:2.1} that the $h\to\gamma\gamma$ matrix element of the operator $O_2(z)$ is proportional to the light-cone distribution amplitude of the photon, and that this quantity approaches the asymptotic form (\ref{phiasy}) when evolved to a high renormalization scale. As long as this matrix element vanishes at the endpoints $z=0$ and $z=1$, the integral in the second term in (\ref{fact1}) is convergent and well defined. If an analogous statement holds for the product of the renormalized jet and soft functions in the third term, such that they tame the behavior of the integrand as $\ell_\pm\to\infty$, then also this integral would be well behaved. These observations motivate us to wonder whether it might be possible to derive a {\em renormalized\/} version of the factorization theorem, which is structurally equivalent to (\ref{fact1}), i.e.\ 
\begin{equation}\label{wonder}
\begin{aligned}
   {\cal M}_b(h\to\gamma\gamma)
   &= H_1(\mu)\,\langle\gamma\gamma|\,O_1(\mu)\,|h\rangle 
    + 2 \int_0^1\!dz\,H_2(z,\mu)\,\langle\gamma\gamma|\,O_2(z,\mu)\,|h\rangle \\
   &\quad\mbox{}+ g_\perp^{\mu\nu} H_3(\mu) \int_0^\infty\!\frac{d\ell_-}{\ell_-} 
    \int_0^\infty\!\frac{d\ell_+}{\ell_+}\,J(M_h\ell_-,\mu)\,J(-M_h\ell_+,\mu)\,S(\ell_+\ell_-,\mu) \,.
\end{aligned}
\end{equation}
Choosing a high value of the renormalization scale, $\mu^2={\cal O}(M_h^2)$, would then cure the endpoint divergences by virtue of resummation. Also, with such a scale choice the first term on the right-hand side would not contain any large logarithms and could be calculated in fixed-order perturbation theory. While it is tantalizing to speculate about this possibility, we have so far not succeeded in deriving a formula such as (\ref{wonder}) from first principles. The bottleneck is that in the regularized theory (before resummation) the various convolution integrals give rise to endpoint divergences, which are regularized either dimensionally (giving rise to $1/\epsilon$ poles) or by means of the rapidity regulator (giving rise to $1/\eta$ poles). In an $\overline{\rm MS}$-like subtraction scheme, one would remove these singularities by applying renormalization factors to the operators and hard functions. In the presence of endpoint singularities, this is however not sufficient to remove all poles. One would thus need to generalize the $\overline{\rm MS}$ scheme and define $Z$ factors for the convolution integrals themselves. The systematics of such an approach will still have to be developed.

\subsubsection*{Acknowledgements}

One of us (M.N.) is grateful to Robert Harlander for suggesting this problem as early as in September 2011, and to Jian Wang for collaborations on related topics at an early stage of this project. We would like to thank Thomas Becher, Guido Bell, Martin Beneke, Philipp B\"oer, Thorsten Feldmann and Ben Pecjak for many useful discussions on endpoint divergences and on the physics behind the refactorization conditions (\ref{refact1}) and (\ref{refact2}). We are especially grateful to Philipp B\"oer for providing us with an electronic version of his PhD thesis, in which endpoint divergences are discussed in the context of exclusive $B$-meson decays. This research has been supported by the Cluster of Excellence PRISMA$^+$\! (project ID 39083149), funded by the German Research Foundation (DFG), and under grant 05H18UMCA1 of the German Federal Ministry for Education and Research (BMBF). The research of Z.L.L.\ is also supported by the U.S.\ Department of Energy under Contract No.~DE-AC52-06NA25396, the LANL/LDRD program and within the framework of the TMD Topical Collaboration.

\newpage
\begin{appendix}
\renewcommand{\theequation}{A.\arabic{equation}}
\setcounter{equation}{0}

\section{Operator matrix elements in SCET-1}
\label{app:A}

In the intermediate effective theory SCET-1, we introduce off-shell, hard-collinear momenta $k_1$ and $k_2$ for the two external photons. They satisfy $-k_i^2={\cal O}(M_h m_b)$, which is parametrically larger than the soft scale $m_b^2$. We can then treat the quark mass as a perturbation and only keep linear terms in $m_b$. At NLO in $\alpha_s$, we obtain for the relevant operator matrix elements 
\begin{equation}
\begin{aligned}
   \langle\gamma\gamma|\,O_1\,|h\rangle
   &= m_{b,0}\,g_\perp^{\mu\nu} \,, \\
   \langle\gamma\gamma|\,O_{2,n_1}(z)\,|h\rangle
   &= \frac{N_c\alpha_{b,0}}{2\pi}\,m_{b,0}\,g_\perp^{\mu\nu}\,
    \frac{e^{\epsilon\gamma_E}\,\Gamma(\epsilon)}{z^\epsilon(1-z)^\epsilon} 
    \left( -k_1^2 \right)^{-\epsilon} \\
   &\quad \times \left[ 1 + \frac{C_F\alpha_{s,0}}{4\pi}\,e^{\epsilon\gamma_E}\,
    \frac{2\Gamma(2\epsilon)\,\Gamma^2(-\epsilon)}{\Gamma(\epsilon)\,\Gamma(3-2\epsilon)}\,
    \Big[ \widetilde K(z) + \widetilde K(1-z) \Big] \left( -k_1^2 \right)^{-\epsilon} \right] , \\
   \langle\gamma\gamma|\,O_3\,|h\rangle
   &= - \frac{N_c\alpha_{b,0}}{\pi}\,m_{b,0}\,g_\perp^{\mu\nu}\,
    e^{\epsilon\gamma_E}\,\Gamma^2(\epsilon)\,\Gamma(1-\epsilon) 
    \left[ \frac{\left( -k_1^2 \right)\left( -k_2^2 \right)}{\left( -M_h^2 \right)} \right]^{-\epsilon} \\
   &\quad\times \left\{ 1 + \frac{C_F\alpha_{s,0}}{4\pi}\,e^{\epsilon\gamma_E} 
    \left[ K_1 \left( \left( -k_1^2 \right)^{-\epsilon} + \left( -k_2^2 \right)^{-\epsilon} \right)
    + K_2 \left[ \frac{\left( -k_1^2 \right)\left( -k_2^2 \right)}{\left( -M_h^2 \right)} \right]^{-\epsilon}
    \right] \right\} ,
\end{aligned}
\end{equation}
where $-k_i^2\equiv-k_i^2-i0$, $-M_h^2\equiv-M_h^2-i0$, and 
\begin{equation}
\begin{aligned}
   \widetilde K(z) &= z^{-\epsilon} \left[ 2(1-\epsilon^2) + 3\epsilon^2(1-\epsilon) z 
    - 2\epsilon(1+2\epsilon) z(1 - z) \right] \\
   &\quad\mbox{}- \frac{\Gamma^2(1+\epsilon)}{\Gamma(1+2\epsilon)} \left[ (1-\epsilon)(2-\epsilon+2\epsilon^2) 
    - 2\epsilon(1+2\epsilon) z(1-z) \right] \\
   &\quad\mbox{}- \frac{2\epsilon}{1-2\epsilon} \left[ 2(1-\epsilon)(1-2\epsilon)
    + \epsilon^2(1+2\epsilon) (1-z) - \epsilon(1+2\epsilon) z(1-z) \right] \\
   &\qquad\times z^{1-\epsilon}\,{}_2F_1(1,1-\epsilon;2-2\epsilon;z) \,, \\
   K_1 &= \frac{3(2-2\epsilon+\epsilon^2)}{\epsilon}\,
    \frac{\Gamma^2(1-\epsilon)\,\Gamma(2\epsilon)}{\Gamma(2-2\epsilon)\,\Gamma(1+\epsilon)} 
    - \frac{2-\epsilon+2\epsilon^2}{\epsilon}\,
    \frac{\Gamma(\epsilon)\,\Gamma^2(1-\epsilon)}{\Gamma(2-2\epsilon)} \,, \\
   K_2 &= \frac{2(1+\epsilon)}{1-2\epsilon}\,
    \frac{\Gamma(1-\epsilon)\,\Gamma^2(2\epsilon)}{\Gamma^2(1+\epsilon)} 
    - 4\,\frac{\Gamma(1-2\epsilon)\,\Gamma^2(2\epsilon)}{\Gamma(1+\epsilon)} \,.
\end{aligned}
\end{equation}
When these expressions are combined with the hard matching coefficients given in (\ref{Hires}), one finds that all endpoint divergences are regularized dimensionally. After renormalizing the bare quark mass and Yukawa coupling using (\ref{myrenorm}), we obtain the finite, off-shell $h\to\gamma\gamma$ amplitude
\begin{equation}\label{Afinres}
   {\cal M}_b(h\to\gamma\gamma) 
   = {\cal M}_0 \left[ \ln\bigg(\frac{-M_h^2-i0}{-k_1^2-i0}\bigg) \ln\bigg(\frac{-M_h^2-i0}{-k_2^2-i0}\bigg) 
    - 2 + \frac{\pi^2}{3} + \frac{C_F\alpha_s}{4\pi} \times \mbox{(finite terms)} \right] .
\end{equation}

\renewcommand{\theequation}{B.\arabic{equation}}
\setcounter{equation}{0}

\section{Exact analytic expression for the soft function}
\label{app:soft}

In (\ref{SaSb}) we have presented some terms in the expression for the soft function using an expansion in powers of $\epsilon$ to the order required for our analysis. We have, however, also obtained closed analytical expressions for the component functions $S_a(w)$ and $S_b(w)$ in terms of generalized hypergeometric functions. We find
\begin{equation}
\begin{aligned}
   S_a(w) &= \frac{e^{\epsilon\gamma_E}}{\Gamma(1-\epsilon)}\,\big( m_{b,0}^2 \big)^{-\epsilon}\,
   r^\epsilon\,(1-r)^{-\epsilon} \\
   &\quad\mbox{}+ \frac{C_F\alpha_{s,0}}{4\pi}\,\big( m_{b,0}^2 \big)^{-2\epsilon}\,e^{2\epsilon\gamma_E}\,
    \Bigg[ \frac{4\Gamma(-\epsilon)\,\Gamma(\epsilon)}{\Gamma(1-2\epsilon)}\,
    r^{2\epsilon}\,{}_2F_1(2\epsilon,1+2\epsilon;1+\epsilon;r) \\
   &\qquad\mbox{}+ \frac{2\Gamma(2-\epsilon)}{\epsilon^3(1-2\epsilon)\,\Gamma(1-3\epsilon)}\,(1-\epsilon r)\,
    r^\epsilon\,(1-r)^{-3\epsilon}\,{}_2F_1(1-2\epsilon,1-2\epsilon;1-3\epsilon;1-r) \\
   &\qquad\mbox{}+ \frac{2\Gamma(-\epsilon)}{\epsilon^2\,\Gamma(1-3\epsilon)}\,r^{1+\epsilon}\,(1-r)^{-3\epsilon}\,
    {}_2F_1(1-2\epsilon,2-2\epsilon;1-3\epsilon;1-r) \\
   &\qquad\mbox{}+ \frac{2\Gamma(\epsilon)}{\epsilon(1-2\epsilon)\,\Gamma(1-\epsilon)}
    \left[ 1 - (1-\epsilon)^2\,(1+2\epsilon)\,r \right] r^\epsilon\,(1-r)^{-1-\epsilon} \Bigg] \,, \\
   S_b(w) &= \frac{C_F\alpha_{s,0}}{4\pi}\,\hat w\,\big( m_{b,0}^2 \big)^{-2\epsilon}\,e^{2\epsilon\gamma_E}\,
    4\Gamma(\epsilon)\,\Gamma(1+\epsilon)\,{}_2F_1(1+\epsilon,1+2\epsilon;2;\hat w) \,.
\end{aligned}
\end{equation}
As previously we use the variables $r=m_{b,0}^2/w$ and $\hat w=w/m_{b,0}^2$.

\renewcommand{\theequation}{C.\arabic{equation}}
\setcounter{equation}{0}

\section{NLO coefficients in the bare decay amplitude}
\label{app:B}

Here we list our explicit findings for the coefficients $k_i$ accounting for the NLO corrections in (\ref{HiOires}). We obtain
\begin{equation}
\begin{aligned}
   k_0 &= - \frac{3}{\epsilon^3} + \frac{1}{\epsilon^2} \left( 2 L_h + 4 L_m - 3 \right) 
    + \frac{1}{\epsilon} \left( - L_h^2 - 2 L_h L_m - 3 L_m^2 + 4 - \frac{\pi^2}{2} \right) \\ 
   &\quad\mbox{}+ \frac13\,L_h^3 + L_h^2 L_m + L_h L_m^2 + \frac53\,L_m^3 + 3 L_m^2 + 2 L_h - (4-\pi^2)\,L_m 
    - \frac{\pi^2}{3} + 8\zeta_3 \,, \\
   k_1 &= - \frac{1}{2\epsilon^4} + \frac{1}{\epsilon^3} \left( L_h - \frac92 \right)
    + \frac{1}{\epsilon^2} \left( - L_h^2 + 3 L_h - \frac{5\pi^2}{12} \right) 
    + \frac{1}{\epsilon} \left( \frac23\,L_h^3 + \frac{5\pi^2}{6}\,L_h 
    - \frac{\pi^2}{4} - \frac{29}{3}\,\zeta_3 \right) \\
   &\quad\mbox{}- \frac13\,L_h^4 - L_h^3 - \frac{5\pi^2}{6}\,L_h^2 
    + \left( 12 + \pi^2 + \frac{58}{3}\,\zeta_3 \right) L_h - 36 - \pi^2 - 5\zeta_3 - \frac{3\pi^4}{16} \,, \\
   k_2 &= - \frac{2}{\epsilon^4} + \frac{4 L_h}{\epsilon^3}
    + \frac{1}{\epsilon^2} \left( - 2 L_h^2 - 4 L_h L_m + 2 L_m^2 + 1 + \frac{2\pi^2}{3} \right) \\
   &\quad\mbox{}+ \frac{1}{\epsilon} \left[ \frac23\,L_h^3 + 2 L_h^2 L_m + 2 L_h L_m^2 - 2 L_m^3
    - L_h \left( 2 + \frac{2\pi^2}{3} \right) - \frac{2\pi^2}{3}\,L_m + 2 + \frac{10}{3}\,\zeta_3 \right] \\
   &\quad\mbox{}- \frac{1}{6}\,L_h^4 - \frac23\,L_h^3 L_m - L_h^2 L_m^2 - \frac23\,L_h L_m^3 + \frac{7}{6}\,L_m^4 
    + \left( 1 + \frac{\pi^2}{3} \right) \left( L_h^2 + 2 L_h L_m \right) \\
   &\quad\mbox{}- \left( 1 - \frac{\pi^2}{3} \right) L_m^2 
    - \left( 4 + \frac{44}{3}\,\zeta_3 \right) L_h + 8\zeta_3\,L_m
    + 4 - \frac{\pi^2}{6} + 8\zeta_3 + \frac{19\pi^4}{60} \,, \\
   k_3 &= \frac{5}{2\epsilon^4}
    +\frac{1}{\epsilon^3} \left( - 2 L_h - 3 L_m + \frac92 \right)
    + \frac{1}{\epsilon^2} \left( L_h^2 + 2 L_h L_m + 2 L_m^2 - 3 L_m - 1 - \frac{\pi^2}{4} \right) \\
   &\quad\mbox{}+ \frac{1}{\epsilon} \left[ - \frac13\,L_h^3 - L_h^2 L_m - L_h L_m^2 - L_m^3
    - \left( 2 - \frac{\pi^2}{3} \right) L_h + \left( 4 + \frac{\pi^2}{6} \right) L_m 
  - 2 + \frac{\pi^2}{4} + \frac{19}{3}\,\zeta_3 \right] \\
   &\quad\mbox{}+ \frac{1}{12}\,L_h^4 + \frac13\,L_h^3 L_m + \frac12\,L_h^2 L_m^2 + \frac13\,L_h L_m^3 
    + \frac{5}{12}\,L_m^4 + L_m^3 + \left( 1 - \frac{\pi^2}{6} \right) \left( L_h^2 + 2 L_h L_m \right) \\
   &\quad\mbox{}- 5 L_m^2 + \left( - 4 + \frac{10}{3}\,\zeta_3 \right) L_h 
    + \left( 8 - \pi^2 - 16\zeta_3 \right) L_m
    - 4 + \frac{7\pi^2}{6} + \zeta_3 - \frac{79\pi^4}{240} \,,
\end{aligned}
\end{equation}
where $L_h=\ln(M_h^2/\mu^2)-i\pi$ and $L_m=\ln[m_b(\mu)^2/\mu^2]$.

\end{appendix}

\end{document}